\documentclass[12pt]{article}
\usepackage[english]{babel}
\usepackage[T1]{fontenc}
\usepackage[a4paper,top=2cm,bottom=2cm,left=3cm,right=3cm,marginparwidth=1.75cm]{geometry}

\usepackage{amsmath}
\usepackage{mathtools}
\usepackage{graphicx}
\usepackage{setspace}
\usepackage{cite}
\usepackage{tikz}
\usepackage{caption}
\usepackage{subcaption}
\usepackage{booktabs}
\usepackage{authblk}
\usepackage{gensymb}
\usepackage{multirow}
\usepackage{csquotes}
\usepackage[colorlinks=true, allcolors=blue]{hyperref}
\usepackage{textcomp, gensymb}
\providecommand{\keywords}[1]
{
  \small	
  \textbf{\textit{Keywords---}} #1
}
\title{Effect of Ti addition on the structural, thermodynamic, and elastic properties of Ti$_x$(HfNbTaZr)$_{(1-x)/4}$ alloys}
\author[1]{Asif Iqbal Bhatti}
\author[1]{Marwa Al-Houcine}
\author[1]{David Tingaud}
\author[1]{Sylvain Queyreau\thanks{Corresponding author: sylvain.queyreau@lspm.cnrs.fr}}

\affil[1]{Université Sorbonne Paris Nord , Laboratoire des Sciences des Procédés et des Matériaux, LSPM, CNRS, UPR 3407, 99 avenue Jean-Baptiste Clément, F-93430 Villetaneuse, France}

\begin{document}
\maketitle

%%%%%%%%%%%%%%%%%%%%%%%%%%%%%%%%%%%%%%%%%%%%%%%%%%%%%%%%%%%%%%%%%%%%%%%%%%%%%%%%%%%
\begin{abstract}
{The structure and thermodynamic properties of Ti$_x$(HfNbTaZr)$_{(1-x)/4}$ from Refractory High Entropy multicomponent Alloys to pure titanium are investigated employing comprehensive MCSQS realizations of the disordered atomic structure and DFT calculations. We showed that to model the random structure in a limited supercell, it is necessary to probe a large space of random configurations with respect to the nearest neighbor’s shells. Mimicking the randomness with the many-body terms does not lead to significant improvements in the mixing energy but modeling the random structure with the few nearest neighbor pairs leads to improvements in the mixing energy. Furthermore, we demonstrated the existence of weak to medium SRO for the two equimolar compositions. Chemical ordering is investigated by associating a large number of MCSQS realizations to DFT energy calculations, and SRO results are rationalized in terms of the crystallographic structure of the pairs of elements and binary phase diagrams. When Ti is added to Ti$_x$(HfNbTaZr)$_{(1-x)/4}$ alloys, the mixing energy remains slightly positive for all $x$. For $x >$ 0.4, a phase transition in favor of an hcp structure is observed in agreement with the predictions of the Bo-Md diagram. At $x$ = 0.5, a dual phase is predicted. Ti content in this class of alloys could be a practical way to select phase structure and tailor the elastic properties to specific applications.}
\end{abstract}

%%%%%%%%%%%%%%%%%%%%%%%%%%%%%%%%%%%%%%%%%%%%%%%%%%%%%%%%%%%%%%%%%%%%%%%%%%%%%%%%%%%

\section{Introduction}
Multicomponent High Entropy Alloys (HEAs), a relatively new class of materials, have recently attracted a great deal of attention due to their remarkable mechanical properties. These HEAs are typically composed of four or more elements at nearly the same concentration \cite{Gao,Miracle2017a,Ikeda2019a}. Despite consisting of a complex multicomponent system, they lead to the formation of a simple single-phase solid solution, such as face-centered cubic (fcc), body-centered cubic (bcc), or hexagonal close-packed (hcp). Among these alloys, fcc \textit{Cantor type} HEAs have been widely studied, but few studies exist on bcc HEAs, and even fewer in the case of bcc refractory HEAs (RHEAs) \cite{Dirras2016,Senkov2018}, sometimes named "Senkov" alloys. The high degree of chemical complexity created by mixing different elements with different atomic sizes and chemical properties increases the material's resistance to deformation and fracture. In addition, the high entropy of the alloy leads to a variety of lattice distortions due to atomic configurations. This arrangement of atoms promotes strain hardening and increases resistance to dislocation movements \cite{Gao2016c,Takeuchi2013,Yao2017,Manzoni2020,Yuan2019}. Despite abundant literature, many aspects of phase stability, exact atomic structure, and their relationship to macroscopic properties remain unclear.

Recent experiments on HfNbTaTi, HfMoTaTiZr, and HfNbTaTiZr have shown that non-toxic and non-allergenic elements such as Nb, Ta, and Zr exhibit high corrosion resistance \cite{Gurel2020}. The latter alloy family, HfNbTaTiZr, will be the focus of the present paper. Properties such as high ductility, corrosion resistance, and strength at both room temperature and high temperatures can be improved by choosing suitable refractory elements (a mixture of bcc and hcp elements).

The d-electron theory has become a classical tool for predicting the structural stability of Ti-based alloys \cite{Abdel-Hady2006} but it has rarely been applied to high-entropy Ti-based alloys. Lilenstein \textit{et al.} have synthesized a novel high entropy alloy (HEA) with the composition Ti$_{35}$Zr$_{27.5}$Hf$_{27.5}$Nb$_{5}$Ta$_{5}$ using the d-electron alloy design approach based on the Bo-Md diagram \cite{Lilensten2017a}. This alloy exhibited a remarkable transformation-induced plasticity effect, resulting in a high normalized work hardening rate while retaining ductility when compared to the reference equimolar Ti$_{20}$Zr$_{20}$Hf$_{20}$Nb$_{20}$Ta$_{20}$ alloy. Microstructure analysis after deformation revealed the large presence of stress-induced orthorhombic martensite. This work shows that the phase stability and mechanical properties of HEA can be tailored by a chemical design approach based on Bo and Md parameters. However, many of the predictions of the Bo-Md diagram need to be compared with the actual HEA phases observed in experiments or simulations.

Departing from the initial picture of the random distribution of atoms, it is now clear that HEA alloys exhibit local chemical fluctuations and various degrees of chemical short-range ordering (SRO), and assessing them is crucial to understanding the microstructure-property relationship. However, characterizing and quantifying composition fluctuations is still a challenging task at the atomic scale, and no direct experimental observation was available until very recently \cite{Lei2018,Bu2021,Zhang2020}. Yin \textit{et al.} were certainly the first to show that Monte Carlo simulations can thus provide an alternative route to study SRO in association with DFT calculations \cite{Yina}, MC was then associated with cluster expansion model \cite{Xun2023} or molecular dynamics simulations \cite{Chen2021}, but there are still only a few studies of this type. An increasing degree of SRO was found to lower the mixing energy of the HEA phase in the corresponding alloys, while some alloys seem more prone to SRO than others. In the quaternary bcc HfNbTiZr alloy, Hf-rich and Ti-rich clusters were experimentally highlighted using atomic probe tomography and HR-HAADF-SEM observations \cite{Lei2018,Bu2021}, in agreement with the affinity of some of the Ti- and Hf-bond types observed in simulations \cite{Zhang2021, Wang2021c}. The electronic density structure analysis performed on the same alloy \cite{Wang2021c} highlighted the peculiar role of Ti-Zr bonds with a reduction of d electrons at the Fermi level, which stabilizes the bcc phase.

Another fascinating aspect of HEA is to fully understand the origin of their large strength, and a number of models have been successful in estimating the yield stress of fcc and bcc alloys \cite{Yin2020, Yin2019, Varvenne2016a}. In deformed HfNbTaTiZr, TEM observations  \cite{Lilensten2018, Couzinie2015} revealed heterogeneous dislocation microstructures, with dense slip bands along \{110\} and \{112\} planes and many debris loops. $\frac{a_{0}}{2}\langle111\rangle$ screw dislocations are mostly rectilinear, and the large apparent activation volumes are consistent with the Peierls mechanism and large friction with the lattice. Once again, the variation of the local chemical environment and SRO existence seem key in rationalizing the dislocation behavior. EXAFS analysis of pure binary and ternary alloys in the Ti-Ta-Hf-Nb-Zr system evidenced local chemical and lattice distortions \cite{Lilensten2022, Tan2023}. MD investigations \cite{Rao2019} of the dislocation structure in the Nb-Ti-Zr system showed core structure variations along the dislocation line depending on the local environment. A collision of cross kinks nucleated on different slip planes yields debris (pairs of interstitial vacancies) in the wake of the dislocation gliding. Yin and collab. \cite{Yin2019} quantified the Peierls barrier for screw dislocations in the bcc MoNbTaW alloy by ab-initio. The dislocation core energy roughly follows a Gaussian distribution depending on the local environment of the dislocation, whose average and variance increase with the degree of SRO. As a consequence, two types of Peierls potentials were identified, and the Peierls barrier was found to increase with the degree of SRO. Finally, in-situ TEM observations revealed \cite{Bu2021} that local chemical fluctuations in the quaternary bcc HfNbTiZr alloy lead to dislocation pinning, stimulating dislocation multiplication and cross-slip activity.

Ti content is expected to strongly impact the stability of the different possible phases and local chemical fluctuations observed in the Hf-Nb-Ta-Ti-Zr system. In this paper, we systematically investigate the effects of Ti content on the atomic structure, phase stability, local chemical ordering, and elastic constants using Monte-Carlo Special Quasi random (MCSQS) and ab-initio calculations. We employ DFT calculations as the derivation of reliable semi-empirical potentials for quinary alloys is still a challenging task \cite{Huang2021}. To better approximate the ideal random structure in a small DFT supercell, we considered multiple atomic structures resulting from different MC SQS calculations, for each of which the energy was calculated. Interestingly, the average mixing energy and Global SRO parameter converge rapidly to constant values with the number of different atomic configurations, while the SRO parameters for each pair type require many more SQS realizations to converge toward the ideal random structure where no pairs are favored. The comprehensive set of data can also be further analyzed to identify favorable, neutral, and unfavorable bond types corresponding to extremes in the mixing energy data. This solution constitutes an alternative to a more complete but more demanding MC-DFT hybrid approach. In agreement with the Bo-Md diagram, tuning the Ti content in HfNbTaTiZr alloys has the potential to alter the alloy's stable phase. The knowledge gained from these investigations can be used to design high-entropy bcc alloys with properties tailored for specific applications. The investigation of the screw dislocation structure and property in correlation to its local chemical environment is left for a forthcoming publication. The remainder of the paper is organized as follows: Section 2 details the simulation conditions and a sensitivity study of the MCSQS parameters to construct the alloy's atomistic configuration. Section 3 provides results and discussion regarding the SRO existing in quaternary and quinary equimolar alloys, the relative stability of the different phases, and the elastic properties as a function of Ti content.  The last section summarizes the conclusions.

%Many of the properties of HEA such as mixing energies or dislocation kinetics, are rooted in the atomic structure of these alloys, which may well deviate from the ideal random structure. However, in the absence of direct experimental observations at the lattice length scale, chemical ordering is mostly unknown and remains a challenging task for simulation approaches.

%%%%%%%%%%%%%%%%%%%%%%%%%%%%%%%%%%%%%%%%%%%%%%%%%%%%%%%%%%%%%%%%%%%%%%%%%%%%%%%%%%%
\section{Methodology and Computational details}

In the present article, we report the mixing energy, elastic properties, and structural parameters of a model of the Ti$_x$(HfNbTaZr)$_{(1-x)/4}$ alloys calculated using Density Functional Theory (DFT) as implemented in Vienna ab-initio simulation (VASP) code \cite{Kresse1996}. The Projector Augmented Wave (PAW) pseudopotential was used to approximate the electron core energy \cite{Kresse1999} and the Perdew-Burke-Ernzerhof (PBE) generalized gradient approximation was used to approximate the exchange-correlation functional (XC) \cite{Perdew1996}.

A lot of care has been paid to the construction of the disordered structure corresponding to the Ti$_x$(HfNbTaZr)$_{(1-x)/4}$ alloys. The composition was varied from $x$ = 0.0 to 0.81 by steps of about 0.1, while the rest of the composition is equally distributed among all the other elements Hf, Nb, Ta, and Zr. For all considered compositions, the supercell geometry was constrained so that the cubic bcc symmetry was preserved, as experiments suggest that most compositions correspond to a bcc single phase. Additional calculations using an hcp, or orthorhombic (ortho) lattice, have also been conducted for high Ti content. Mimicking a random structure in a supercell of sufficiently limited dimensions to be tractable by DFT simulations raises an additional problem due to short-range spatial correlations among atoms induced by the Periodic Boundary Conditions (PBC). To address this problem, several techniques have been proposed in the literature based on the use of order parameters to evaluate the disordered nature of the atomic configuration and minimize the spatial correlations as much as possible \cite{Glass2006,Soven1967,Gao2017b,Song2017,Carlsson1986}. Here, we use the Monte-Carlo Special Quasi-random Structure (MCSQS) technique implemented in the Alloy Theoretic Automated Toolkit (ATAT) code \cite{VandeWalle2013,VanDeWalle2009} to generate and analyze a large set of different random structure realizations. We can thus perform a statistical study of the MCSQS parameters on the resulting structure of different atomic configurations.
%SQ: Celine told me there are two different orthorhombic structures. do you which is which? AI: There is one primitive, and the other is bcc ortho structure. I have modeled in this paper with bcc ortho. => SQ: ok great

The supercell size was chosen to be a necessary tradeoff between physical relevance and the numerical cost of the calculations. The supercell geometries depend on the investigated composition and are as follows: For $x$ = 0.11, 0.33, 0.4, 0.63, and 0.70, we generated a supercell of 54 atoms based on $3\times3\times3$ cubic bcc unit cells; for compositions corresponding to $x$ = 0.0, 0.50, and 0.81, a supercell of 128 atoms based on $4\times4\times4$ cubic bcc unit cells was used; and for $x$ = 0.20, a supercell of 125 atoms based on $5\times5\times5$ primitive bcc unit cells was used. In the case of hcp structures, we employed supercells of 128 atoms for $x$ = 0.5, 0.81 ($4\times4\times4$ hexagonal unit cells), and 54 atoms for $x$ = 0.63, 0.70 ($3\times3\times3$ hexagonal unit cells).

In all the reported simulations, the system was allowed to fully relax, including atomic positions, supercell shapes, and volumes (in an anisotropic fashion), using a conjugate gradient algorithm. Brillouin zone integrations were performed with a $\Gamma$-centered k-point grid scheme. The plane-wave cut-off energy was set to 600 eV. The first order Methfessel-Paxton method \cite{Methfessel1989} with a smearing parameter of 0.2 eV was used. The system was assumed to have converged when the force on each atom was less than 1 meV/\AA, and the stress tolerance was below $~0.005$ GPa. After relaxation, the mixing energy was evaluated using a single-point energy calculation with a dense k-point relative to the convergence of the mixing energy.  A sensitivity study showed that a tolerance of 1 meV/atom, a $2\times2\times2$ k-point grid, and a plane-wave cut-off energy of 600 eV were sufficient for the calculation of elastic and structural properties. 

%%%%%%%%%%%%%%%%%%%%%%%%%%%%%%%%%%%%%%%%%%%%%%%%%%%%%%%%%%%%%%%%%%%%%%%%%%%%%%%%%%%
\section{Results and Discussion}
We employed the MCSQS approach to mimic the random structure of HEA in a DFT supercell. While MCSQS is now a well-established technique within the HEA community, its results can be impacted by a few parameters, in particular, the number of neighbor shells and the order of the many-body terms that are included to evaluate the disorder in the considered configuration. We thus start with a sensitivity study of the structure generated as a function of these parameters.

In the present work, we have varied the pairwise terms (hereafter called \textit{p}) between the NN (Nearest Neighbor) shells, where \textit{p} ranges from 2 to 10 NN. We have also varied the three-body (t) and four-body (q) terms (where t and q range from 0 to 1 NN).

\begin{figure}[ht]
    \centering
    \includegraphics[scale=0.14]{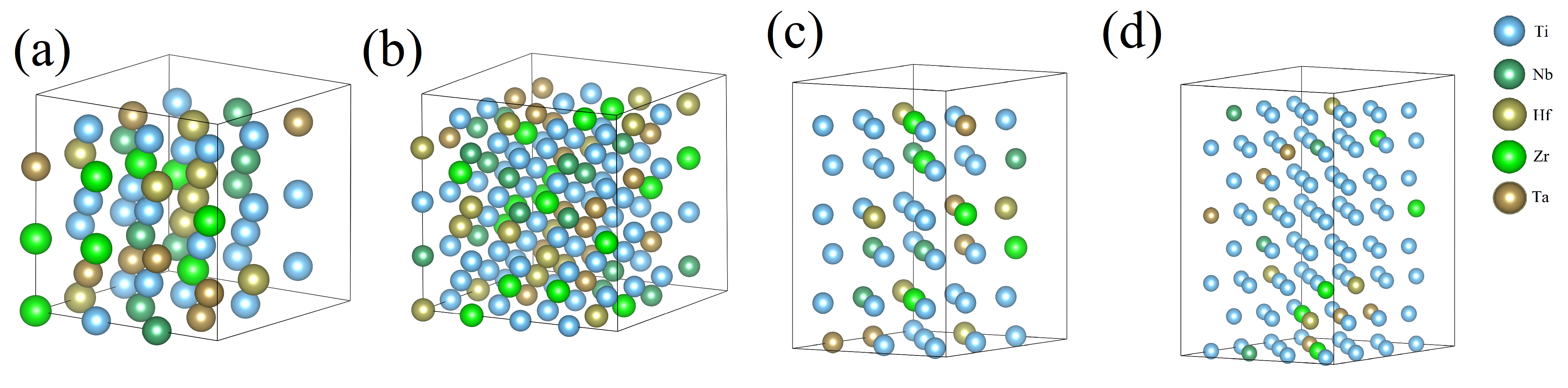}
    \caption{Examples of supercells [(a)-(d)] obtained from MCSQS for the alloy Ti$_x$(HfNbTaZr)$_{(1-x)/4}$. Figures (a) and (c) show a $3\times3\times3$ supercell for $x$ = 0.11, 0.33, 0.40, 0.60, and 0.70 containing 54 atoms with bcc and hcp structures, respectively. Figure (b) and (d) show a $4\times4\times4$ supercell for $x$ = 0.0, 0.50, and 0.80, representing 128 atoms with a bcc and hcp structure, respectively.}
    \label{fig:Figure_1_SQSsupercell}
\end{figure}

For each Ti content $x$ and MCSQS parameters p, t, and q (hereafter denoted by the value triplet (p, t, q)), Figure \ref{fig:Figure_1_SQSsupercell} shows some of the atomic configurations resulting from the MCSQS calculations obtained for different Ti content $x$. Our workflow consists of first generating 20 SQS for (p, 0, 0) only, then with (p, t, 0), and finally (p, t, q). In each case, the MCSQS calculation was initiated with different random seeds, leading to different final atomic configurations. The configuration was assumed to be final when the local minimum of the objective function was reached during the 120-hour runtime. MCSQS was terminated earlier when no further SQS structure was generated for more than 12 hours.

According to classical theory, the Gibbs free energy of the system is given by:

\begin{equation}\label{eq:Gibbs1}
    \centering
    \Delta G_{x} = \Delta H_{x} - (T\Delta S_{mix})_{x} - (T\Delta S_{vib})_{x}
\end{equation}

In Eq. \ref{eq:Gibbs1}, $\Delta$H$_{x}$ is the mixing energy, and $\Delta$S$_{mix}$ and $\Delta$S$_{vib}$ are the mixing and vibrational entropy terms, respectively. The mixing energy, $\Delta$H$_{x}$, given in Eq. \ref{eq:Gibbs2} is computed from the difference of energy of the SQS supercell (E$_{SQS}$) and the energy E$_{i}$ associated with the most stable phase of each element, i (bcc for Nb, and Ta and hcp for Hf, Ti, and Zr): %SQ: Celine told me that when not using the stable pure element as we do latter, this is called mixing energy? I did have time to check but it could simplify the explanation of Figure 8 %AI: what I saw people use mixing enthalpy. => we cannot say formation energy when using natural (hcp and bcc) structure and mixing energy when using forced bcc structure for all then? too bad

\begin{equation}\label{eq:Gibbs2}
    \centering
    \Delta H_{x} = (E_{SQS})_{x} - (\sum_{i}c_{i}E_{i})_{x}
\end{equation}

where c$_{i}$ is the atomic fraction of an alloy element, \textit{i}. For alloys, the mixing entropic term is as follows:

\begin{equation}\label{eq:Gibbs3}
    \centering
    (\Delta S_{mix})_{x} = - k_{B} \sum_{i}c_{i}ln(c_{i})
\end{equation}

where $k_{B}$ is the Boltzmann constant. It can be noted that Eq. \ref{eq:Gibbs3} applies to perfect random solid solutions, which might not be the case for HEA alloys, where some level of chemical order may exist (\textit{see later}). Eq. \ref{eq:Gibbs3} is nevertheless commonly used within the HEA community to estimate the mixing entropy. The last term of Eq. \ref{eq:Gibbs1}, corresponding to the vibrational entropy, can be calculated from DFT using the so-called frozen phonon approach \cite{Alfe2009, Togo2015}. However, these types of calculations are computationally demanding, and vibrational entropy is commonly neglected when compared to mixing entropy \cite{Du2019, Sobieraj2020,Song2017,Gao}. This last term is thus disregarded in the present study.

%%%%%%%%%%%%%%%%%%%%%%%%%%%%%%%%%%%%%%%%%%%%%%%%%%%%%%%%%%%%%%%%%%%%%%%%%%%%%%%%%%%
\section{Sensitivity study on MCSQS parameters}

First, we conducted MCSQS calculations while neglecting the many-body terms, i.e., we included only the pairwise \textit{p} NN shells, as these terms are expected to have the largest impact on the resulting SQS configuration. We then determined the effect of the number of p shells on the mixing energy. To evaluate this parameter, we have considered the $x = 0.33$ composition. Table \ref{tab:Formation_H} shows the average mixing energy and average lattice parameter as a function of p shells. The values given in Table \ref{tab:Formation_H} refer to the lowest mixing energy obtained over a relevant number of different atomic configurations. We will see later that 20 different SQS realizations are sufficient to assess the ideal solid solution mixing energy for the alloys and supercell sizes considered here. Interestingly, the mixing energy per atom decreases only weakly as the number of p NN shells increases up to 8 NN. For $p$ = 10 NN, it surprisingly yielded a larger mixing energy per atom by about 7 meV/atom. In the latter case, it is probable that modeling SQS supercells with 10 pairs of shells within a 120-h time limit is not enough, or it might be that considering a larger number of shells within a finite supercell constrained by the given composition and size of the supercell overlaps due to periodic boundary conditions (PBC). As a result of this sensitivity study, we chose $p$ = 8 NN as a compromise between MCSQS capacity to produce low energy configurations and the numerical cost to run the MCSQS.

\begin{table}[ht]
    \centering
    \caption{Calculated average lattice parameters a$'$, and mixing energy, $\Delta$H, for $x$ = 0.33 using a 54-atom supercell over 20 different initial SQS seeds (i) for each value of p along with the average (avg) mixing energy value and standard deviation ($\sigma$). The number of nearest neighbor (NN) shells is given by the following convention: p NN where p is the number of nearest neighbor shells for the pair.}
    \label{tab:Formation_H}
    \begin{tabular}{cccccc}
        \toprule
         \#Shells && min\{$\Delta$H\} & avg $\pm$ $\sigma$ && a$'$ \\ 
         \cmidrule{3-4}\cmidrule{6-6}
         (p,0,0) NN && \multicolumn{2}{c}{meV/atom} && \multicolumn{1}{c}{\AA} \\
         \midrule
         2  && 84.73 & 88.5$\pm$3.2 && 3.382 \\ 
         4  && 82.37 & 89.0$\pm$3.9 && 3.384 \\ 
         6  && 81.27 & 89.7$\pm$3.6 && 3.383 \\ 
         8  && 80.37 & 86.8$\pm$3.8 && 3.383 \\ 
         10 && 87.30 & 90.5$\pm$2.6 && 3.382 \\ 
         \bottomrule
    \end{tabular}
\end{table}

\begin{figure}[ht]
    \centering
    \includegraphics[scale=0.7]{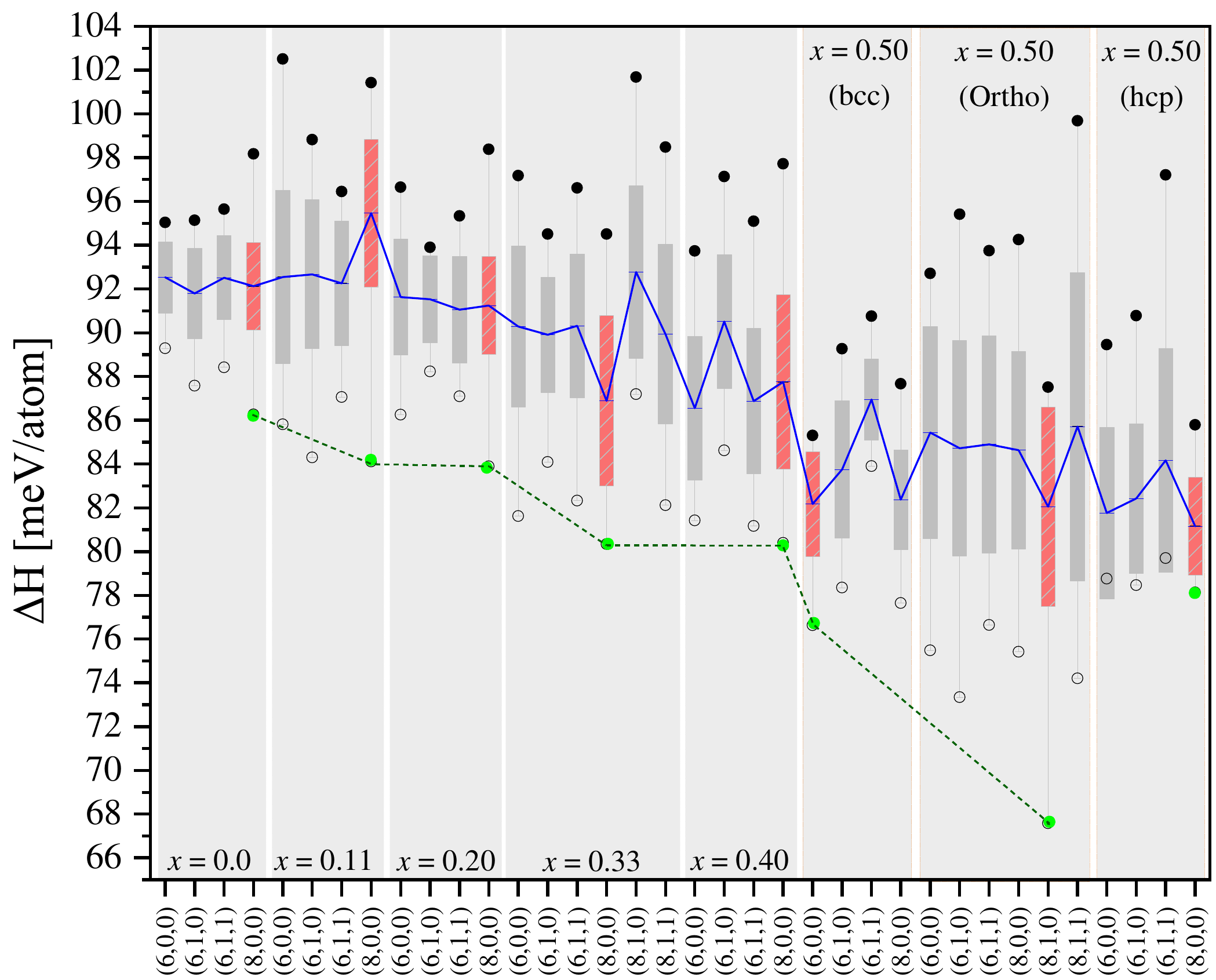}
    \caption{Statistical plot showing the average mixing energy, minimal and maximal values, and the standard deviation of the calculated mixing energy for Ti$_x$(HfNbTaZr)$_{(1-x)/4}$ alloys. The number of shells is given by the following convention: (p, t, q), where p, t, and q are the pair, three-body, and four-body terms, respectively. Maximum values are represented by black-filled circles, and minimum values by green circles. The average mixing energy for each composition is shown in a blue dotted color line, and the standard deviation is shown in a gray rectangle. For each $x$ = 0.0 and 0.20, 650 SQS calculations were performed, respectively.}
    \label{fig:SQS_statistics}
\end{figure}

In another set of calculations, we have investigated the possible effect of higher order many-body terms: ternary (t) and quaternary (q) interactions within 1st NN shells from MCSQS. For this, many-body terms were set to $t$ = 1 NN, and $q$ = 0 or 1 NN, while the pairwise terms were set to 6 and 8 NN shells. Figure \ref{fig:SQS_statistics} shows the impact of including these higher order many-body terms on the mixing energy of Ti$_x$(HfNbTaZr)$_{(1-x)/4}$ for $x$ ranging from 0.0 to 0.5. The figure reports the average (avg) mixing energy value, along with the standard deviation ($\sigma$) and minimum and maximum energy values. Overall, average mixing energy decreases as Ti content $x$ increases. Further, the average mixing energy over 20 initial seeds is weakly impacted when three and four-body terms are included. The standard deviation is relatively large for some of the composition and simulation conditions. This could be the result of the impact of favorable and unfavorable pair types (see below); it could also be due to the lower mechanical stability of the bcc phase in the presence of the hcp prone elements. 

For a given Ti content $x$, the average mixing energy (shown in a blue dotted color line) decreases only weakly when including t and q terms. A similar conclusion was also reached by Tian \textit{et al.} for CoCrFeMnNi, Ta-W, and CoCrNi alloys. Using the Similar Atomic Environment (SAE) approach, these authors showed that including many-body clusters did not provide much improvement in the cross-validation error between SAE and SQS for equimolar composition, implying that pair NN terms for atom pairs are relatively enough to model a random structure \cite{Tian2020a}.

%AI: I checked my data and there was a shift in values, which I have corrected now for x=0.11 (8,0,0).

At $x$ = 0.5, near the extrapolated red line (see Figure \ref{fig:Bo_Md_pred}) we could have a phase transition; therefore, we generated another 20 SQS structures with orthorhombic and hcp symmetry, respectively. In this case, the orthorhombic structure generated with (8, 1, 0) is 7 meV/atom less than in other cases shown in Figure \ref{fig:SQS_statistics}. It shows that for low-symmetry structures, the (t, q) might play a role in modeling the un-equimolar composition. For the hcp structure, we see the trend in mixing energy for all (p, t, q) is almost the same. However, there seems to be competition between bcc and hcp with an energy difference of 2 meV/atom. Thus, we conclude that at $x$ = 0.50, hcp $>$ bcc $>$ ortho. Among all the configurations, we selected the one associated with the lowest mixing energy.

The addition of three and four body terms up to 1 NN shell appears to have only a second-order effect in the present system. The inclusion of a large number of NN shells for pair interactions has an important impact on the ability of the SQS to yield lower energy configurations. This sensitivity study of over 700 different calculations demonstrates the importance of performing multiple MCSQS calculations with different initial seeds to evaluate the statistical differences that result from the MCSQS calculations.

%%%%%%%%%%%%%%%%%%%%%%%%%%%%%%%%%%%%%%%%%%%%%%%%%%%%%%%%%%%%%%%%%%%%%%%%%%%%%%%%%%%
Next, we focus on the quaternary and quinary systems. The atomic configuration obtained from the MC SQS technique constitutes only one of the realizations of the system, and the alloys and sizes considered here only represent a part of the many atomic arrangements seen in a random solution. Thus, different MC SQS realizations could lead to different configurations and thus provide a more complete picture of the structure of the considered systems. We chose the equimolar composition, i.e., $x$ = 0.0 and 0.20, for which we generated a large set of 650 distinct SQS structures using the optimized SQS parameters ($p = 8$, $t = 0$, $q = 0$ NN pairs). SQS calculations were run to produce a random structure for maximizing the objective function using the above mentioned convergence criteria. After this, the relaxation of each structure was done using DFT, and the mixing energy was evaluated for all different initial configurations.  The data can thus be analyzed to see how many SQS calculations are needed to provide a realistic picture of the ideal random solid solution. For this, we will consider two aspects: i) the average mixing energy, and ii) the SRO parameter (see below).

Figure \ref{fig:TE_vs_SQS} displays the energy per atom $E_{at}^{i}$ and associated scatter obtained from each of our MC SQS runs 's', taken in the order they were created. The figure also shows the running average of the energy per atom: $\bar{E_{at}}(k) = \sum_{s=1}^k E_{at}^{s}/k $. Interestingly, this latter quantity converges rapidly to a roughly constant average of $\bar{E_{at}}(k)$ after about 20 MC SQS realizations for both equimolar alloys. This convergence of the average energy is obviously very empirical here, as sorting the raw energy data in a different order might change the running average evolution but not its final value. For the moment, we will simply retain the trends in the evolution of quantities and the number of MC SQS realizations required for average convergence.

To quantify the SRO in these simulations, we use the SRO parameter $\alpha_{ij}$ described by Fontaine \textit{et al.}  \cite{DeFontaine1971},

\begin{equation}\label{eq:Fontaine}
    \centering
    \alpha_{ij} = \frac{\frac{n_{j}}{m_{A}} - c_{j}} {\delta_{ij} - c_{j}}
\end{equation}

Here, $i$ is the reference atom. $n_{j}$ is the number of atoms of the non-reference type among the $c_{j}$ atoms in the $i$-th coordination shell. $m_{A}$ is the concentration of the non-reference atom. If $i = j$, $\delta_{ij} = 1$; otherwise, the Warren-Cowley SRO parameter is returned. The total SRO, $\alpha = \sum_{ij} |\alpha_{ij}|$, is calculated by summing all pairwise terms around the reference atom in the bcc structure for the first two and three nearest neighbor coordination shells (2 NN and 3 NN), respectively. We employed the tool developed in \cite{Menon2019} and extended it to perform the analysis, including the first three nearest neighbor coordination shells (3 NN).

\begin{figure}[hp]
    \captionsetup[subfigure]{labelformat=empty}
    \centering
    \begin{subfigure}[b]{0.85\textwidth}  
        \includegraphics[width=\textwidth]{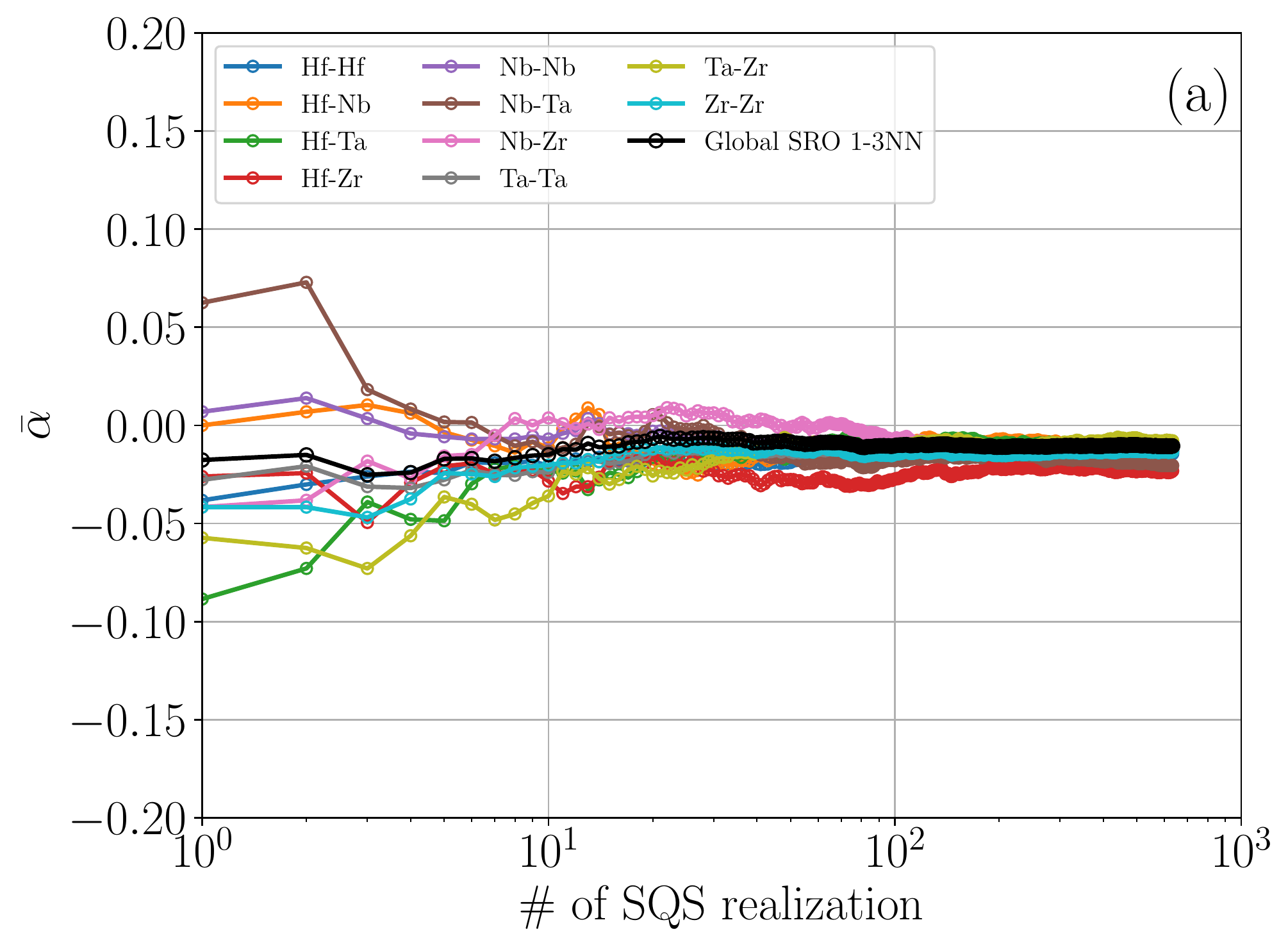}
        %\caption{$~$}
    \end{subfigure}
    \hfill
    \centering
    \begin{subfigure}[b]{0.85\textwidth} 
        \includegraphics[width=\textwidth]{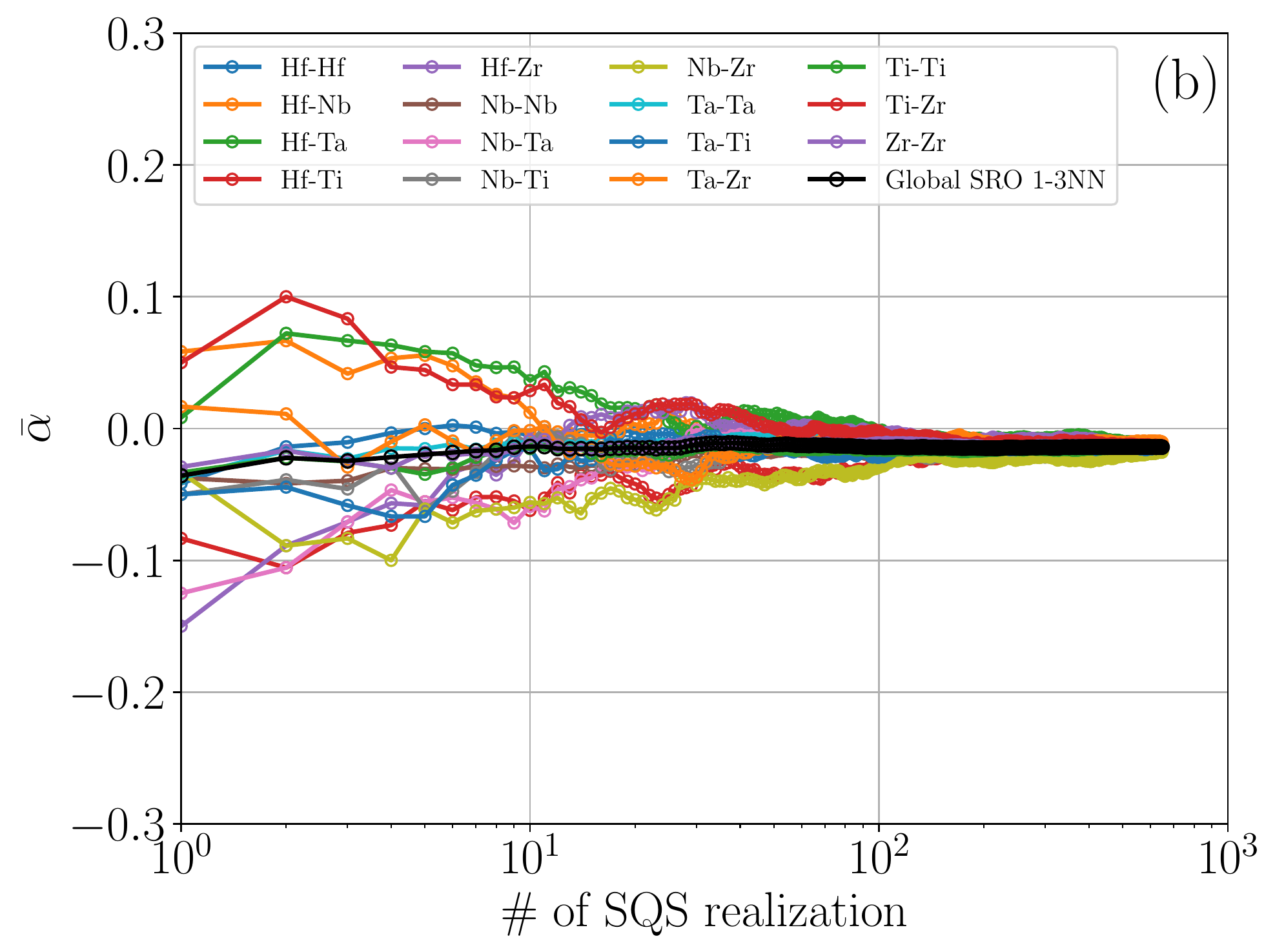}
        %\caption{$~$}
    \end{subfigure}
    \caption{Running averages of the pairwise $\alpha_{ij}$ and Total SRO parameters $\bar{\alpha}(k) = \sum_{s=1}^k \alpha^s/k/N_{ij} $ as a function of the growing number of SQS realizations included in the average, for (a) $x$ = 0.0 and in (b) $x$ = 0.20.}
    \label{fig:SRO_vs_SQS}
\end{figure}

Figure \ref{fig:SRO_vs_SQS} shows the evolution of the total SRO parameter $\bar{\alpha}(k) = \sum_{s=1}^k \alpha^s/k/N_{ij} $ as a function of the considered SQS realizations included in the running average, where $N_{ij}$ is the number of bond types. First, it can be noted that the SRO values remain rather small. The total SRO parameter (black curve) converges rapidly to a somewhat constant value when considering more than 20 different MC SQS realizations. %SQ: I don't think that I used the same global SRO parameter as you since I am not using the absolute sign, the figure looks better for this particular section. %AI: For this, you have used only the 3NN not the sum of the first three shells? Is that right? => SQ: the sum 1 to 3 NN shells
The final global SRO value is not perfectly equal to zero as expected for the ideal random solid solution, but is negligibly close with a value of -0.01. The SRO analysis shown in Figure \ref{fig:SRO_vs_SQS} was done on the 3NN shells. Interestingly, the results of the SRO analyses on the 1NN and 2NN shells showed similar trends, but running average convergence was achieved for a smaller number of MC SQS realizations.

The figure also shows the running averages of the SRO parameters $\alpha_{ij}(k)$ corresponding to each bond type. Quite expected, these quantities require a larger number of MC SQS (between 20 and 100) before converging toward the ideal random solid solution ($\alpha_{ij} \to 0$). Conversely, the empirical convergence analysis carried out here highlights the need to perform and consider several MC SQS realizations, at least for the alloys and simulation conditions used here. If a similar variety of configurations were sought in a single simulation box, this would represent sizes one or two orders of magnitude larger than what has been considered in the present study, and such simulation sizes are currently mostly out of reach for DFT calculations.

\begin{figure}[hp]
    \captionsetup[subfigure]{labelformat=empty}
    \centering
    \begin{subfigure}[b]{0.9\textwidth}  
        \includegraphics[width=\textwidth]{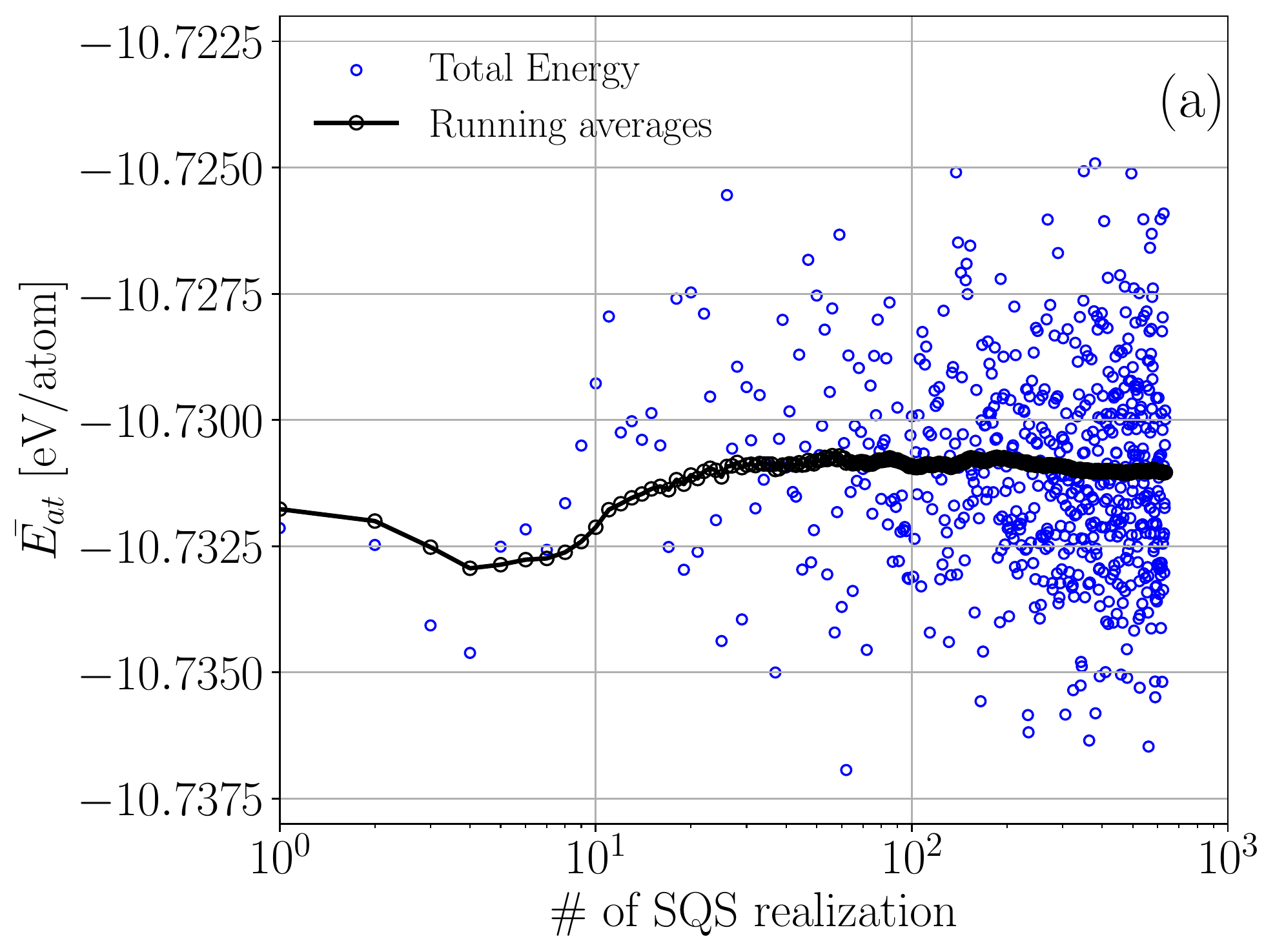}
        %\caption{$~$}
    \end{subfigure}
    \hfill
    \centering
    \begin{subfigure}[b]{0.9\textwidth} 
        \includegraphics[width=\textwidth]{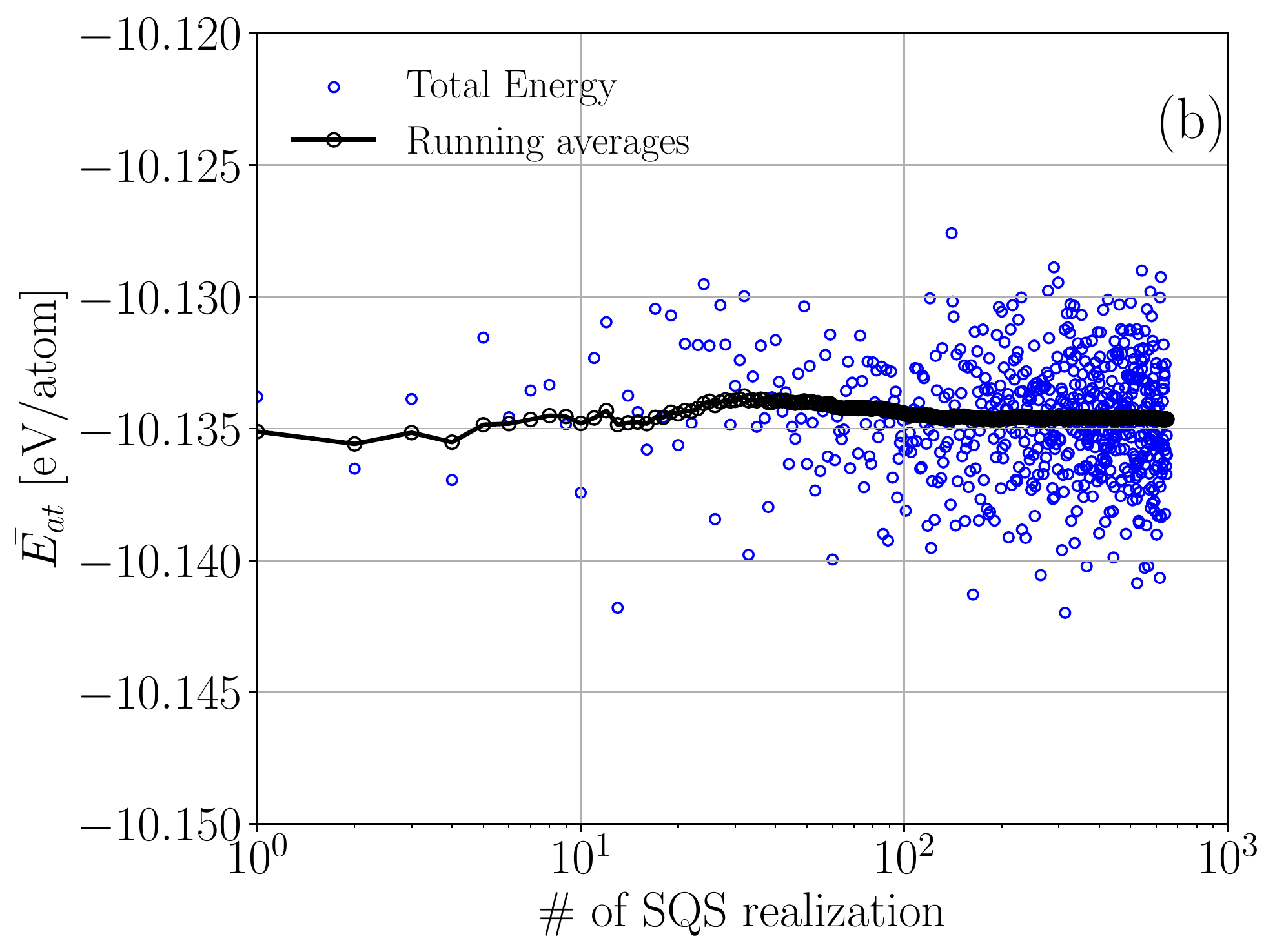}
        %\caption{$~$}
    \end{subfigure}
    \caption{The calculated total energy per atom for every 650 distinct SQS structures is shown in blue, while the running average, $\bar{E_{at}}(k) = \sum_{s=1}^k E_{at}^{s}/k $ as a function of the growing number \textit{k} of SQS realizations is shown with black line for (a) $x$ = 0.0 and in (b) $x$ = 0.20.}
    \label{fig:TE_vs_SQS}
\end{figure}

%%%%%%%%%%%%%%%%%%%%%%%%%%%%%%%%%%%%%%%%%%%%%%%%%%%%%%%%%%%%%%%%%%%%%%%%%%%%%%%%%%%
\section{Analysis of the chemical short-range order of the bcc phase}

The results of the previous section raise the question of whether the statistical fluctuation could be understood among all SQS structures that exhibit a local CSRO, i.e., whether the favorable pairings between elements are preferred, leading to a configuration with the lowest mixing energy. MC and DFT calculations have been associated recently to investigate the short-range order (SRO) in two bcc alloys \cite{Yin2020,Zhang2021}, MoNbTaW and HfNbTiZr, respectively. The development of SRO was shown to decrease the mixing energy of the solid solution phase. Here, we show that the large number of different SQS simulations allows us to sample the energy of a disordered atomic configuration in a less computationally demanding way. We continue focusing on the two equimolar alloys and employing the 650 SQS structures obtained from the optimized SQS parameters ($p = 8$, $t = 0$, $q = 0$ NN pairs).  In Figure \ref{fig:SQS_statistics}, for $x$ = 0.0 and 0.20, the low and high mixing energies of the configurations are shown, along with the standard deviation and average values.

To assess the SRO, we sort the different atomic structures with respect to the simulated energy configuration. Following the lines of previous sections, we checked the significance of this sorting by analyzing the average of all 650 different configurations. Thus, the first set of data corresponds to the average of all the data and is shown as green curves in Figures \ref{fig:Tix0_LE}, \ref{fig:Tix0_bcc} and \ref{fig:Tix0.2_LE}. As desired, the SRO parameter is close to zero for all pair types. A small deviation on the SRO parameter of 0.01 is found on this average, and this gives an estimate of the precision we can expect from our analysis. The fact that the SRO parameter is very close to zero means that the average of all 650 SQS configurations is a good approximation of the ideal random solid solution. This neutral curve, representing an ideal solid solution, will act as a reference level line for the next steps. Two bins were then considered: the configurations with the lowest energy are grouped and called the low energy (LE) set, while the configurations with the highest energy are called the high energy (HE) set. To determine the SRO for the LE SQS set, we averaged the SRO values of each pair type for a group of "LE" cases with the lowest energy and used the same strategy for the HE SQS set. We varied the size of these bins from 6 to 65 configurations ($1-10\%$ of the total data), and a larger bin size smooths the results by reducing the obtained averaged SRO parameter values. However, the sign of the SRO parameters, indicating whether a bond type is more or less present than in the ideal random structure, remained largely unchanged when the bin size was varied.

\subsection{SRO in the quaternary HfNbTaZr system}

Before comparing the results of our two compositions, Figures \ref{fig:Tix0_LE} and \ref{fig:Tix0.2_LE} show the radar plot for $x$ = 0.0 and $x$ = 0.2, and the analysis is admittedly not straightforward. This analysis was performed for the first 1-3 NN shells, where most of the SRO can be observed. An SRO parameter with a negative sign indicates a pair type that is more abundant in the atomic structure under consideration than in the ideal random configuration. When comparing two different SQS realizations, an increase in mixing energy can be explained in two ways: a lower representation of favorable pair types and/or an increase in the contribution of unfavorable atomic pairs. Comparing the results for the LE and HE sets, we see that the SRO data for the two sets agree qualitatively well, i.e., favorable bonds are found in greater numbers in the LE sets ($\alpha_{ij} < 0$) and unfavorable bond types in smaller numbers ($a_{ij} > 0$). However, the trend is less clear when the SRO parameter is close to zero, indicating a neutral influence of the corresponding pair.

Turning now to the specific interactions in quaternary alloys, among the 10 different bond types, the following pairs appear to be favorable (in decreasing order): Nb-Ta and Hf-Zr; the same atom pairs (e.g., Hf-Hf) show a slight interaction ($\alpha_{ij} < 0$); finally, the last pair types seem to be unfavorable (in decreasing order): Hf-Nb, Ta-Zr, Nb-Zr, and finally Hf-Ta. Interestingly, the classification of a bond "ij" can also be determined in a completely different way, namely by a linear fit of the configurational energy as a function of the parameter set $a_{ij}$ calculated for all 650 SQS realizations, where favorable (or unfavorable) bonds are associated with a negative slope (positive) of the linear fit. This further strengthens confidence in the proposed SRO analysis.

Previous results strikingly indicate a simple correlation between the favorable nature of the bond types and the crystallographic reference structure of the two elements involved in the pair type. This is shown in Figure \ref{fig:Tix0_bcc}(a), where an SRO parameter is defined as the sum of SRO values corresponding to pairs of elements depending on their structure: bcc-bcc, hcp-hcp, or bcc-hcp elements (at 0 K). While this analysis is an oversimplification of the more complex picture in Figure  \ref{fig:Tix0_LE}, elements with identical reference crystallographic structures typically seem to lead to favorable bonds associated with a decrease in mixing energy, while those with different bcc-hcp crystallographic structures are unfavorable in this system. It would be interesting to nuance this rather simple sorting of SRO analysis by the fact that the elements here have a similar radius and chemical properties precisely chosen to promote the formation of an HEA, namely a unique solid solution. If we had bcc elements other than the elements in the quaternary alloys, the findings would be much more complex. What is more, it can also be noticed that the binary phase diagrams of the elements involved in the favorable pairs lead to a unique solid solution, regardless of composition, while the elements of the unfavorable pairs give rise to a mixture of binary solid solutions. Figure \ref{fig:Tix0_LE}(b) shows the SRO parameter $a_{ij}$ when performing the analysis of an increasing number of shells from one to three NN shells for the LE SQS realizations. Interestingly, the first NN shell is associated with rather weak SRO parameters for all bond types, so the chemical SRO is mostly present in the 2nd and 3rd NN shells, which have a larger number of atomic layers. 
%[SQ]In this system, the SRO parameter values for the 2nd and 3rd NN shells are mostly the same, indicating that most of the SRO occurs in the 2nd NN shell.
%\textcolor{red}{[SQ?] mixture of binary solid solutions}

\subsection{SRO in HfNbTaTiZr system and comparison between the two systems}

Some of the previous observations made for the quaternary alloy can be repeated here, although the details may differ. Figure \ref{fig:Tix0.2_LE}(a) shows that the average SRO parameters for all 15 bond types are close to zero when considering all 650 SQS realizations, consistent with the ideal random solid solution. The LE and HE data point sets again complement each other well, with some exceptions that we will discuss later. The following bond types appear to be uniquely favorable in this new system ($\alpha_{ij} < 0$ for the LE set): Hf-Zr, Hf-Ti, and Nb-Ta; the same element pairs "i-i" are typically neutral ($\alpha_{ij} = 0$) or weakly favored, while the following bonds appear unfavorable ($\alpha_{ij} > 0$): Hf-Ta, Nb-Ti, and Ta-Zr. Once more, the binary phase diagram of Hf-Ti, the additional favorable pair compared to the quaternary, shows one solid solution, whatever the chemical composition. The results for some bond pairs, such as Ta-Ti and Ti-Ti, are inconclusive with our methodology. More SQS realizations are certainly needed to analyze this complex quinary system.

%\textcolor{red}{[SQ:?] Once more, the binary phase diagram of Hf-Ti, the additional favorable pair compared to the quaternary, shows one solid solution, whatever the chemical composition.} 

\begin{figure}[hp]
    \captionsetup[subfigure]{labelformat=empty}
    \centering
    \begin{subfigure}[b]{0.45\textwidth} 
        \includegraphics[width=\textwidth]{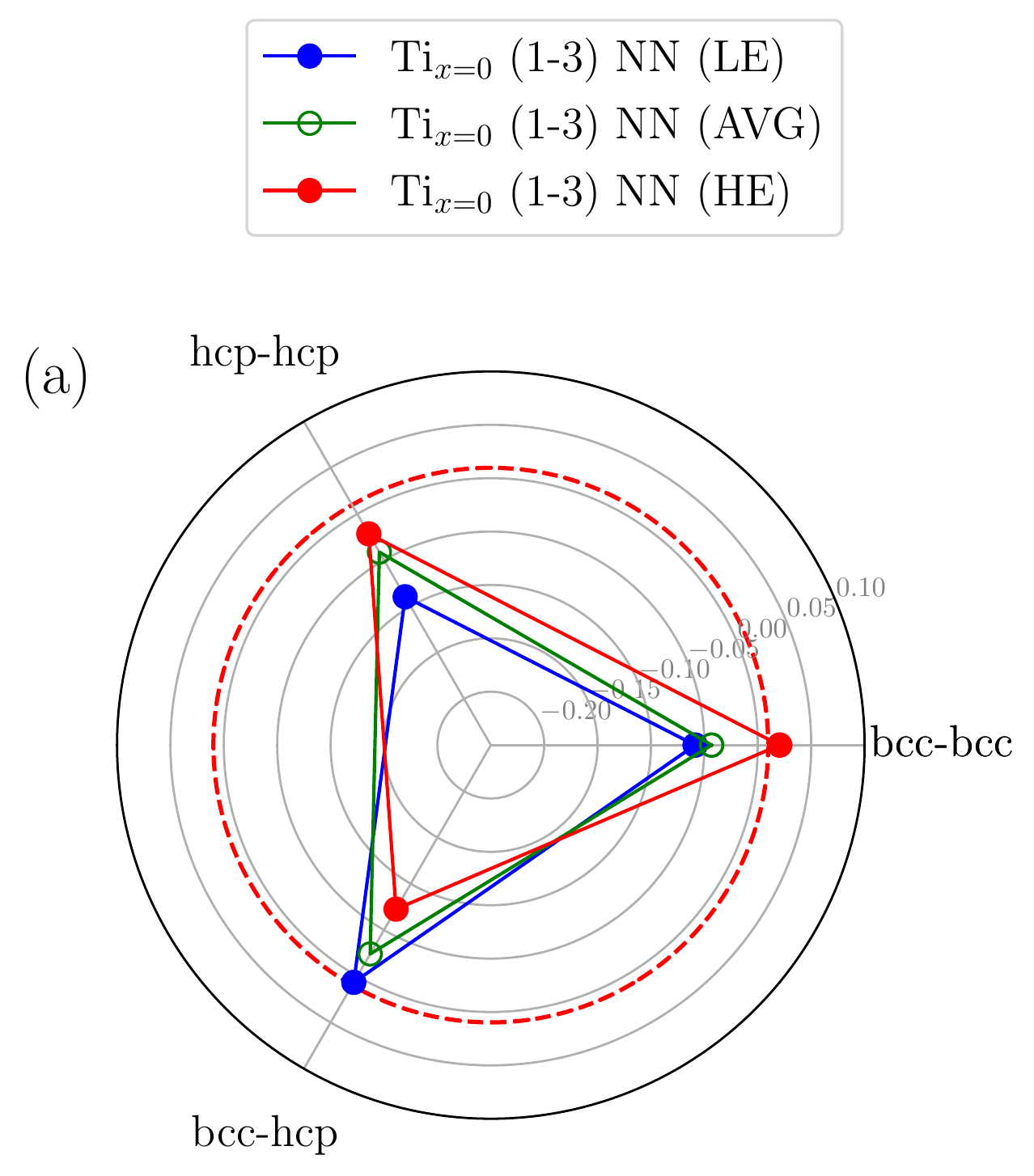}
        %\caption{$ $}
    \end{subfigure}
    %\hfill
    \centering
    \begin{subfigure}[b]{0.45\textwidth} 
        \includegraphics[width=\textwidth]{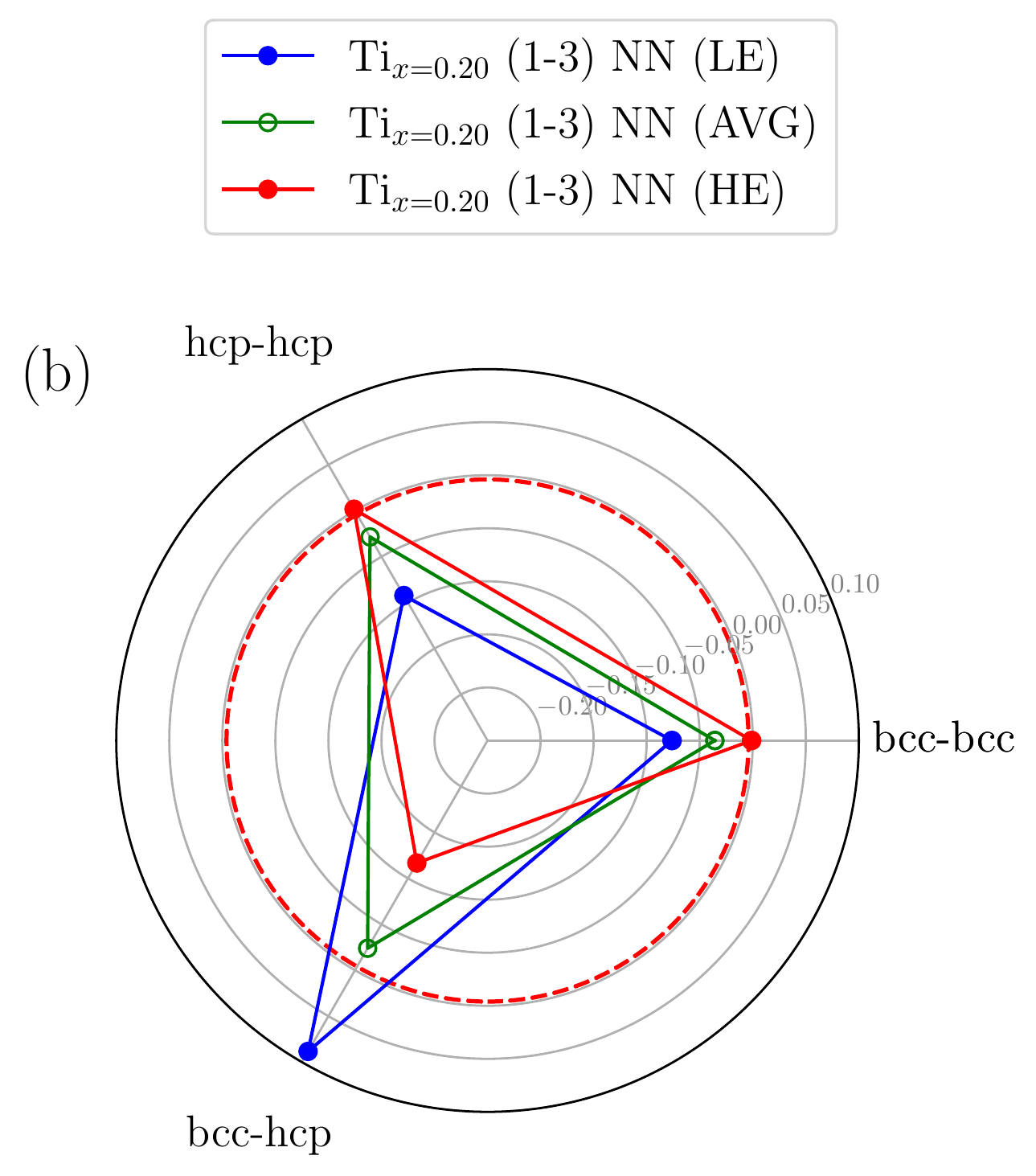}
        %\caption{$ $}
    \end{subfigure}
    \caption{Radar plot showing the average, the low energy (LE) set and the high energy (HE) set averaged over fifteen data points for the case 1st + 2nd + 3rd (1-3) NN data points for (a) $x = 0$ and (b) $x = 0.20$, respectively. The LE and HE SQS sets are averaged over fifteen data points when the data are regrouped into "simple pair categories", "bcc-bcc", "bcc-hcp" and "hcp-hcp" based on the stable phase for the individual atoms of the pair. "AVG" here refers to the average over all SQS realizations. The red dashed line represents the ideal random solid solution.}
    \label{fig:Tix0_bcc}
\end{figure}

\begin{figure}[hp]
    \captionsetup[subfigure]{labelformat=empty}
    \centering
    \begin{subfigure}[b]{0.45\textwidth}  
        \includegraphics[width=\textwidth]{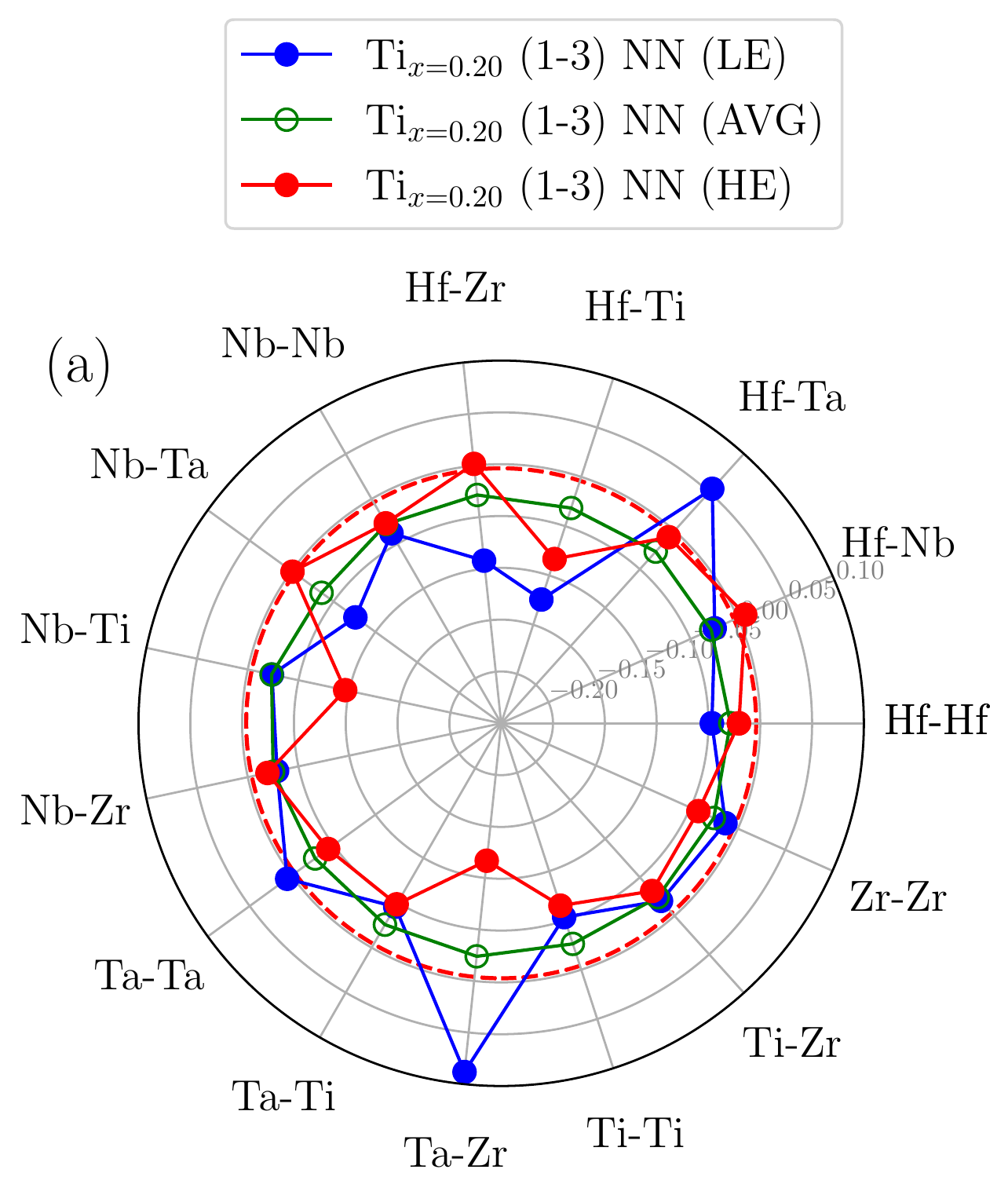}
        %\caption{$ $}
    \end{subfigure}
    %\hfill
    \centering
    \begin{subfigure}[b]{0.45\textwidth} 
        \includegraphics[width=\textwidth]{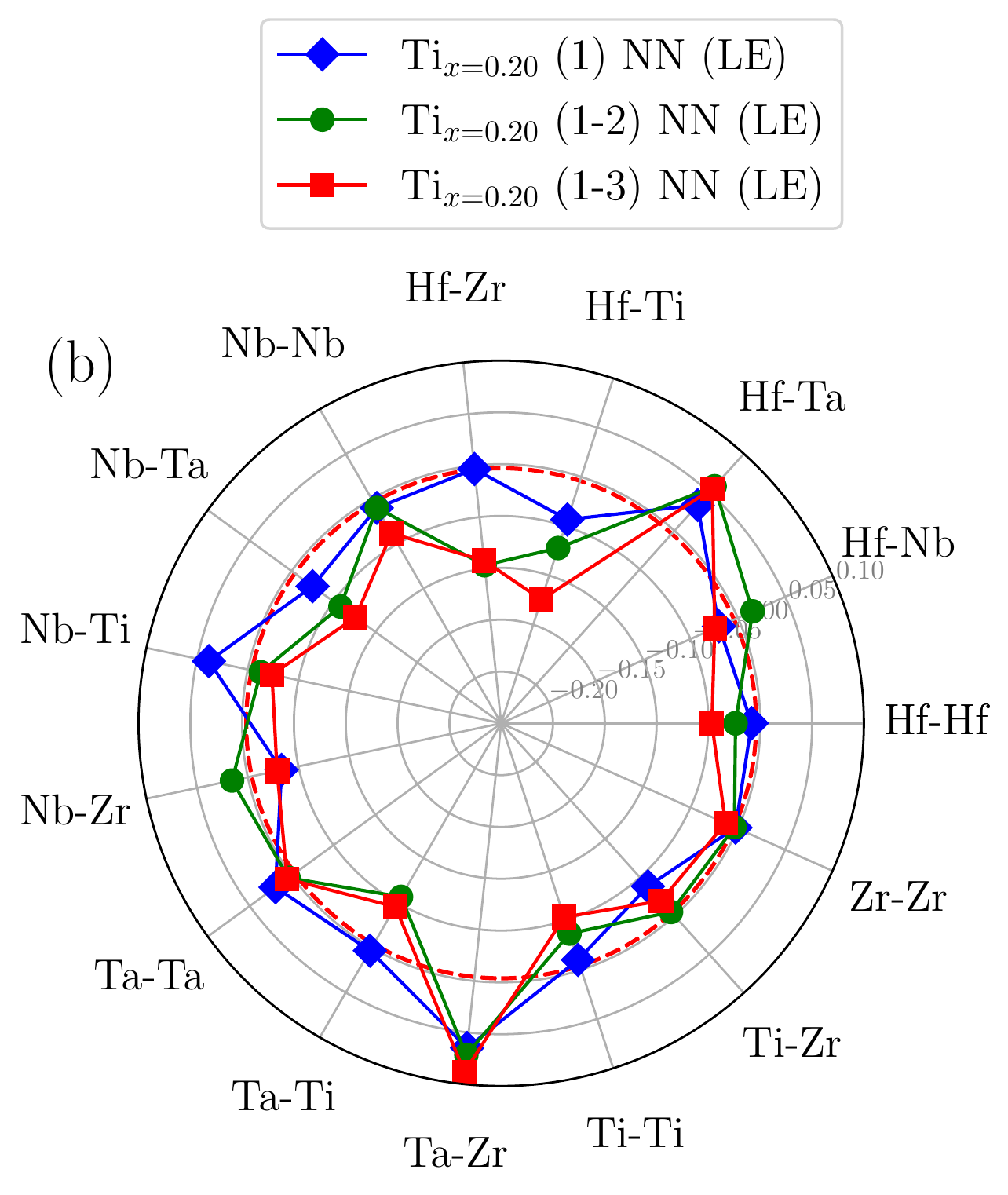}
        %\caption{$ $}
    \end{subfigure}
    \caption{Idem as Figure \ref{fig:Tix0_bcc}, but for $x = 0.20$.}
    \label{fig:Tix0.2_LE}
\end{figure}

The simple sorting of bond types as a function of the crystallographic structure of each element is shown in Figure \ref{fig:Tix0_bcc}(b). As with the quaternary alloy, the sum of the SRO parameters is favorable for pairs of elements with the same bcc-bcc or hcp-hcp structures, although the SRO parameter for hcp-hcp has a lower value here. The SRO parameter for bcc-hcp element bonds is unfavorable and has a much larger value than in the quaternary system. Finally, the SRO parameters for all 15 pair types are analyzed as a function of NN shells and shown in Figure \ref{fig:Tix0.2_LE}(b). In the quinary system, the bond types are polarized even in the 1st NN shell, where the SRO value closely follows the sorting by the crystallographic nature of the elements involved in the bond (the agreement is better than when considering 1st + 2nd NN shells). The SRO data for other shells typically evolves uniformly from this 1st NN shell to the following 2nd and 3rd NN shells. Thus, the chemical order in the quinary system appears to be more complex and more spread than in the quaternary systems, as the 1st, 2nd, and 3rd NN shells are involved.

\begin{figure}[ht]
    \captionsetup[subfigure]{labelformat=empty}
    \centering
    \begin{subfigure}[b]{0.45\textwidth}  
        \includegraphics[width=\textwidth]{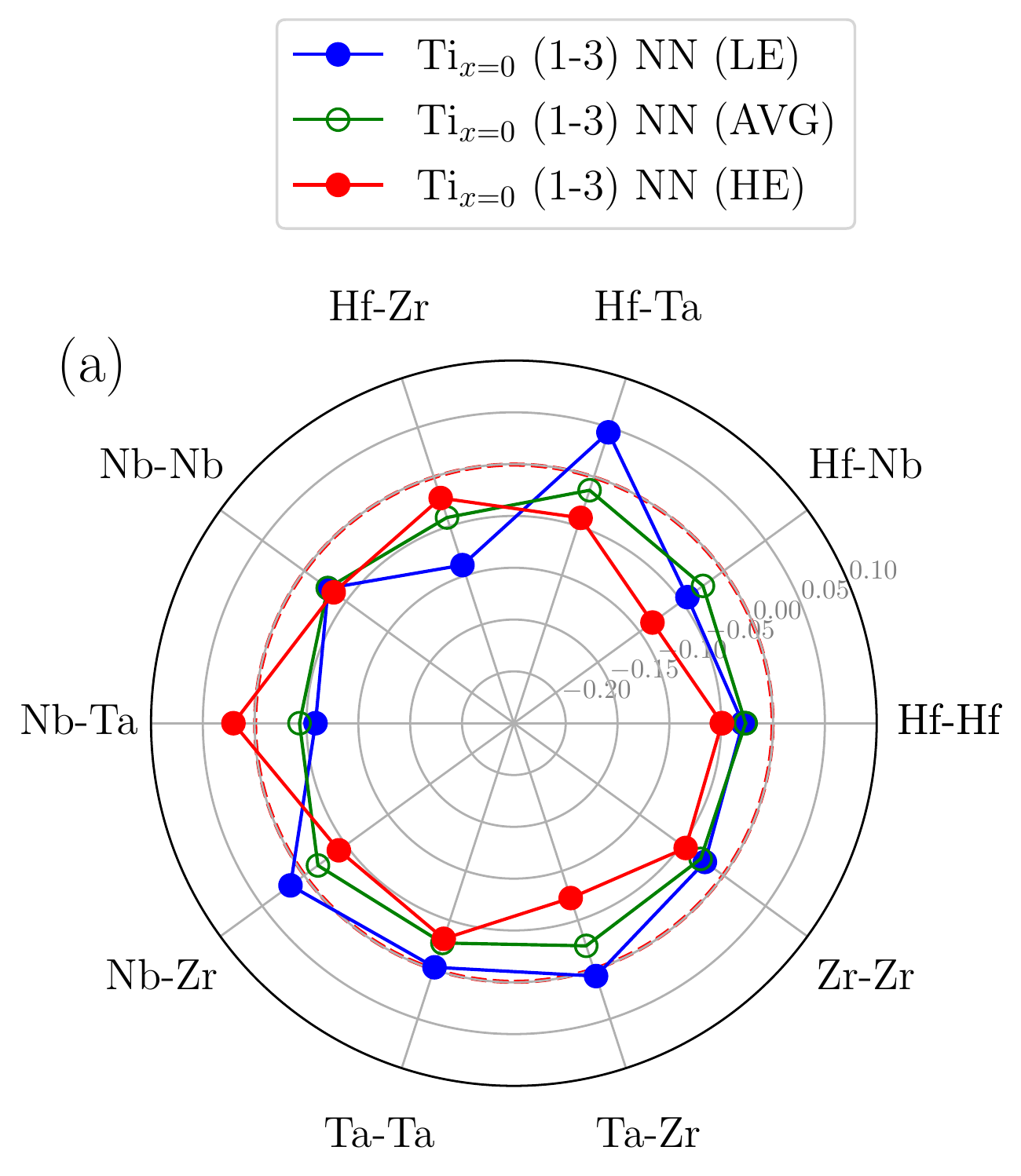}
        %\caption{$~$}
    \end{subfigure}
    %\hfill
    \centering
    \begin{subfigure}[b]{0.45\textwidth} 
        \includegraphics[width=\textwidth]{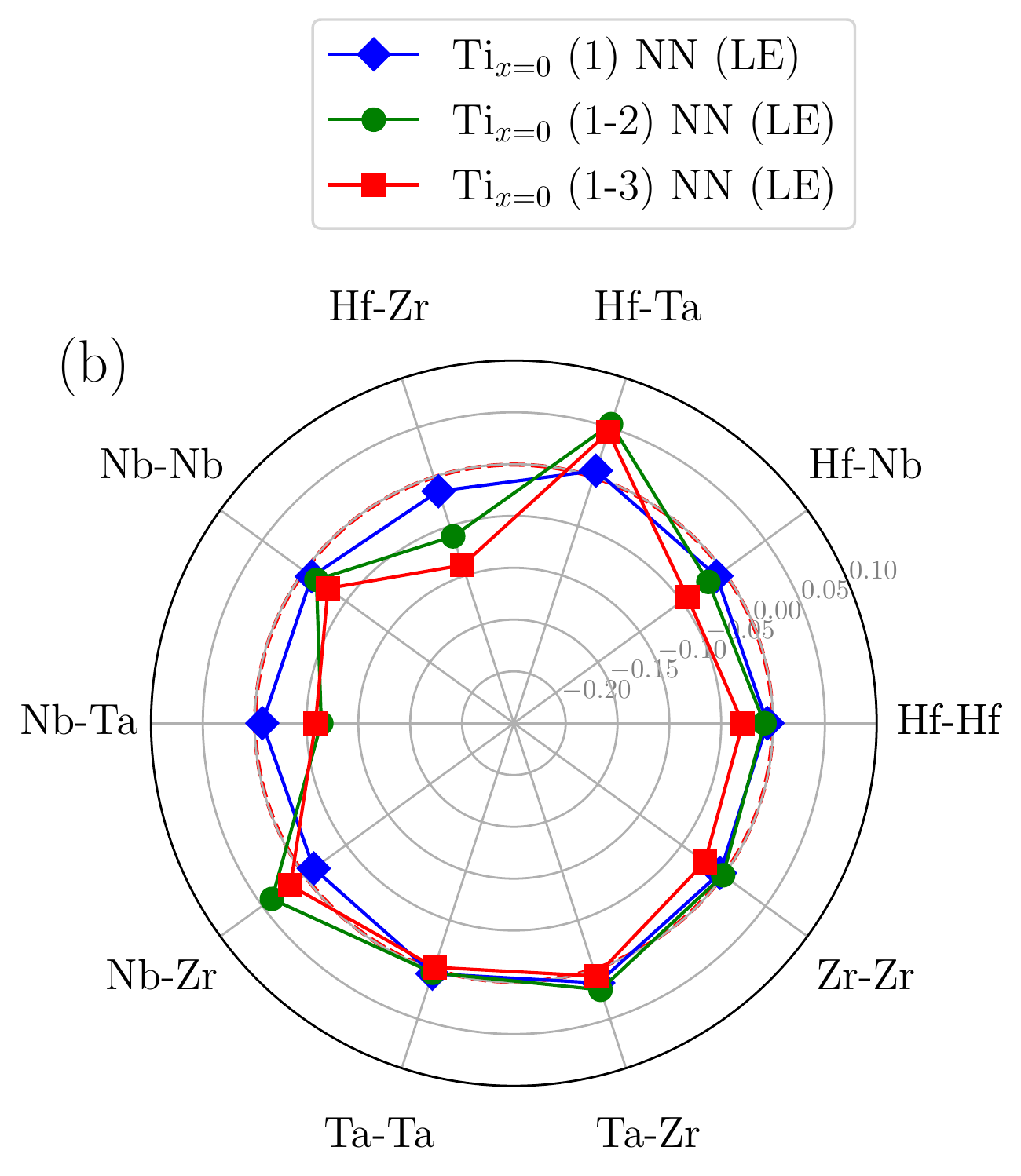}
        %\caption{$~$}
    \end{subfigure}
    \hfill
    \caption{Radar plot showing the average values for $x$ = 0: low energy (LE) set averaged over fifteen data points, and high energy (HE) set averaged over fifteen data points, for (a) 1st + 2nd + 3rd (1-3) NN data points and (b) low energy (LE) set SQS realizations for the 1st (1) NN, 1st + 2nd (1-2) NN, and 1st + 2nd + 3rd (1-3) NN shells, respectively. The LE and HE SQS sets are an average of the first and last fifteen data points, respectively. "AVG" here refers to the average over all SQS realizations. The red dashed line represents the ideal random solid solution.}
    \label{fig:Tix0_LE}
\end{figure}

The total SROs for $x$ = 0.0 and 0.20 are given in Table \ref{tab:SRO_value} and are calculated from the sum of the first two and the sum of the first three NN shells. In our analysis, we find that for $x$ = 0.0, we have the lowest SRO value for the lowest energy configuration compared to the highest energy configuration. For $x$ = 0.20, the SRO values are higher for the configurations with the lowest energy, indicating the existence of SRO. When comparing the quaternary and quinary alloys, the SRO in the first system is mostly localized on the 2nd NN shell, and only 2 bond types (Nb-Ta and Hf-Zr) are favorable, 4 bond types are unfavorable, and the same element bonds are mostly neutral. In the quinary system, 4 bond types are favorable, 6 types are unfavorable, and 5 types are slightly less than zero, while the chemical ordering covers at least all three NN shells. Since, in a finite-sized supercell, the bond type contributions are related through the imposed overall alloy composition, the quaternary system seems more constrained, and this could explain why the total SRO is smaller in that system. In the study performed by Yin \textit{et al.} using the Monte Carlo approach (MC), the author observed a large Warren-Cowley SRO value (in the range of $1.75-2.00$) for bcc MoNbTaW RHEA for the lowest energy configuration by considering the first two NN shells \cite{Yin2021}. However, it should be noted that the quaternary system they studied is arguably simpler as it consists of only bcc elements. %For alloys consisting of bcc and hcp elements, the local ordering appears to be complex.

\begin{table}[ht]
    \centering
    \caption{Total SRO of the lowest SQS configuration (LSQS) and highest SQS configuration (HSQS) for the combined 1st+2nd nearest neighbors (represented as 2 NN) and 1st+2nd+3rd nearest neighbors (represented as 3 NN) shells, respectively. Total SRO is determined by using SRO = $\sum_{ij} |\alpha_{ij}|$. \#atoms indicates the number of atoms in a SQS supercell.}
    \label{tab:SRO_value}
    \begin{tabular}{ccccccc}
        \toprule
         Ti$_x$(HfNbTaZr)$_{(1-x)/4}$&\#atoms& 2NN& 3NN& & 2NN& 3NN \\ 
         \cmidrule{3-4}\cmidrule{6-7}
         $x$ && \multicolumn{2}{c}{LSQS} && \multicolumn{2}{c}{HSQS} \\ 
         \midrule
          0.0&128&	0.57&	0.51&&	0.88&	0.81\\
          0.2&125&	1.55&	1.61&&	1.46&	1.48\\
         \bottomrule
    \end{tabular}
\end{table}

Our results regarding the chemical SRO are mostly in qualitative agreement with the limited existing literature, and in particular with another very recent MC investigation on the same quinary alloys \cite{Xun2023}. The MC relied on a cluster expansion model fitted on DFT calculations and conducted at different finite temperatures. While the values of the WC SRO parameter are quantitatively different (and larger) for the various pair types, the authors concluded that Hf-Ti and Zr-Ti showed the strongest chemical preference, followed by Hf-Nb, Zr-Nb, Nb-Ta, and finally Ta-Ta. The SRO was analyzed at small (1st NN only) and larger length scales (1st to 4th NN), and the SRO appeared weaker when analyzed at the latter scale. The strong affinity among Ti-Zr, Ti-Hf and Nb-Hf elements was also observed in the similar HfNbTiZr bcc alloy from both simulation \cite{Zhang2021,Wang2021c} and experimental investigations \cite{Lei2018,Bu2021}. In experiments, Hf-rich and Ti-rich clusters were identified from HR-HAADF-SEM and APT  measurements \cite{Lei2018,Bu2021}. The enthalpy of mixing for Ti-Zr and Hf-Nb atomic pairs close to zero or negative was suggested as an explanation \cite{Xun2023, Wang2021c}. The electronic density structure analysis in \cite{Wang2021c} highlighted the role of Ti-Zr SRO in particular, with a reduction of d electrons at the Fermi level, which was found to stabilize the bcc phase. 

Similarly to our findings, the quaternary HfNbTaZr alloy seemed to be less prone to local chemical ordering in \cite{Xun2023} when compared to other similar Ti-based HEA alloys, suggesting a particular role of Ti atoms in the SRO development. Our SRO analysis in Figure \ref{fig:Tix0_LE}(a) matches very well with the results in \cite{Xun2023} on the same alloy (for data from 1st to 4th NN).

%%%%%%%%%%%%%%%%%%%%%%%%%%%%%%%%%%%%%%%%%%%%%%%%%%%%%%%%%%%%%%%%%%%%%%%%%%%%%%%%%%%
\section{Possible phases according to the Bo-Md diagram}

To predict the phase stability of our alloys, we reprise the \textit{d}-electrons theory initially proposed to predict the phase stability of Ti-based alloys. Morinaga \textit{et al.} originally proposed the theory to classify Ti-based alloys into $\alpha$ or $\alpha + \beta$ or $\gamma$ phase as a function of solely two parameters: Bo and Md \cite{Morinaga1988}. Bo is the bond strength between the Ti and the alloying elements, and Md is the metal's d-orbital energy level, which corresponds to the element's metallic radius and electronegativity \cite{Morinaga1988}. Based on this, Hadi \textit{et al.} extended the single $\beta$-phase field to the domain from Bo $\leq$ 2.84 to Bo $\leq$ 2.96 \cite{Abdel-Hady2006}.

For Ti-based alloys, Bo and Md parameters are given as composition averages: Bo = $\sum$ c$_{i}$Bo$_{i}$ and Md = $\sum$ c$_{i}$Md$_{i}$. The $c_{i}$ is the concentration of the alloying elements. Table \ref{tab:Bo_Md} lists the composition averages for the alloys considered here. The reference Bo and Md values for each element in bcc Ti are taken from Hadi \textit{et al.}  \cite{Abdel-Hady2006}. We will assume that this analysis is still valid for our alloys, in which Ti is not the main constituent in some of our nuances  \cite{Lilensten2017a}.

\begin{table}[ht]
    \centering
    \caption{The Bo and Md for the Ti$_x$(HfNbTaZr)$_{(1-x)/4}$ alloys are calculated using the reference values of Bo and Md from Hadi \textit{et al.} \cite{Abdel-Hady2006}.}
    \label{tab:Bo_Md}
    \begin{tabular}{cccccccccc}
        \toprule
         $x$& 0.0& 0.12& 0.20& 0.33& 0.40& 0.50& 0.63& 0.70& 0.811\\
         \midrule
         Bo&3.110&	3.074&	3.046&	3.003&	2.979&	2.95&	2.908&	2.884&	2.850\\
         Md&2.715&	2.685&	2.662&	2.626&	2.607&	2.582&	2.548&	2.528&	2.499\\
         \bottomrule
    \end{tabular}
\end{table}
%\textcolor{red}{$\overline{Bo}-\overline{Md}$ diagram [SQ: sometimes w/o overline on Bo and Md when we mention the diagram, we need to pick notation] AI: I have followed the Lilensten approach where he uses without overline. The average is understood from the context. }

\begin{figure}[ht]
    \centering
    \includegraphics[scale=0.65]{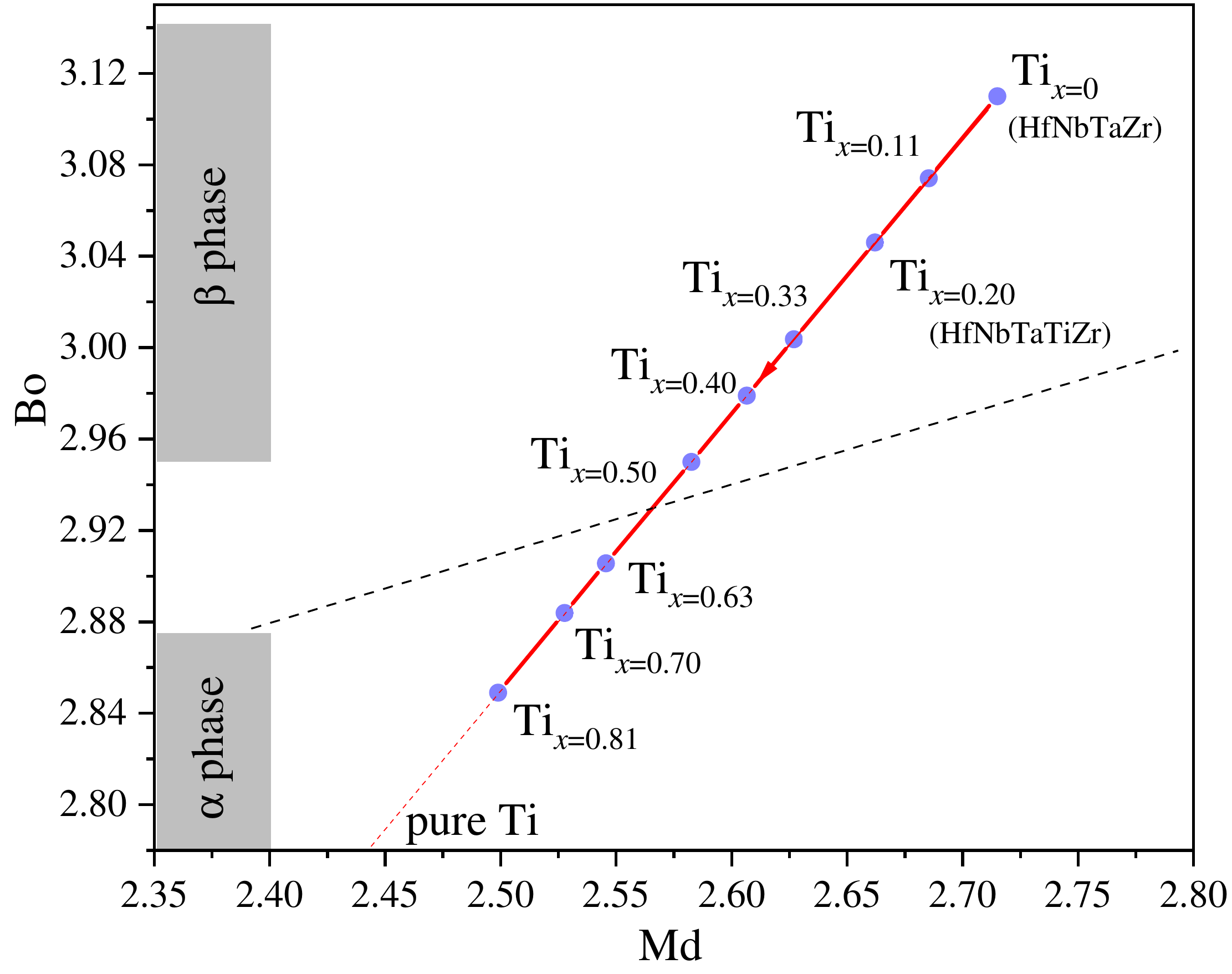}
    \caption{Extrapolation of the Bo-Md diagram for $Ti_{x}$ compositions. Designed alloys are shown with a solid red line. The dashed black line shows the separation between the beta and alpha phases obtained from the reference \cite{Abdel-Hady2006}.}
    \label{fig:Bo_Md_pred}
\end{figure}

In the Bo-Md diagram shown in Figure \ref{fig:Bo_Md_pred}, the domain of stability associated with the various possible phases of Ti based alloys are drawn along with some examples from the literature \cite{Abdel-Hady2006,Lilensten2017a}. The solid line in red is extrapolated for Ti$_x$(HfNbTaZr)$_{(1-x)/4}$ alloys with the composition studied in the present work to pure Ti system. For compositions well above the dashed line in the Bo-Md diagram, a bcc structure is expected, while for compositions well below the same line, structures tend to become hcp. For $x$ = 0.0 (the quaternary alloy without Ti) and $x$ = 0.2 (the equimolar case), experimental studies exist and provide clear reference data in the shape of unambiguous identification of a single HEA phase of bcc structure \cite{Huang2021,Senkov2011}, corresponding to well-defined X-ray diffraction peaks for the bcc structure. The Bo-Md diagram predictions are thus in agreement with these studies on HEA alloys. From this, it becomes clear that relevant MCSQS and DFT simulations must be provided not only for the bcc phase but also for the hcp phase at larger Ti content $x$. 

%%%%%%%%%%%%%%%%%%%%%%%%%%%%%%%%%%%%%%%%%%%%%%%%%%%%%%%%%%%
\section{Structure and stability of the bcc phase as a function of the Ti content}

In this section, we present DFT results on the effects of Ti content on the structure and relative stability of the possible phases. For the remainder of the paper, simulation results are presented for the average of 20 MCSQS runs of the best LE sets.

\begin{figure}[ht]
    \centering
    \includegraphics[scale=0.30]{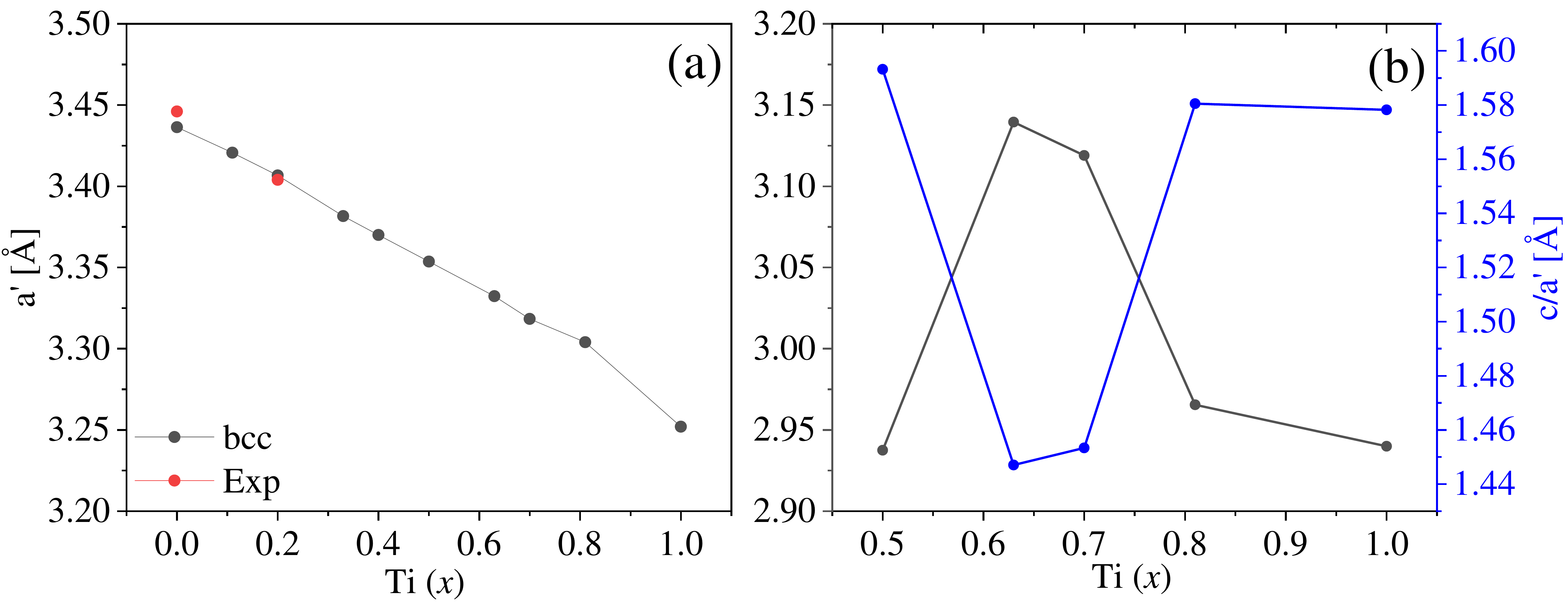}
    \caption{Calculated average lattice parameters a$'$ as a function of Ti content $x$ are reported for the conventional unit cell. In (a), the average lattice parameters for the bcc phase are shown, and in (b), the average lattice parameters for the hcp phase are shown. The red dots are the experimental lattice parameters for the given composition \cite{Huang2021, Senkov2011}. The tabulated data can be found in the supplementary section (see Table \ref{tab:S1}).}
    \label{fig:lattice}
\end{figure}

In Figure \ref{fig:lattice}, the average lattice parameter (a$'$ = (a+b+c)/3 for bcc) decreases from 3.436 \AA ~to 3.252 \AA ~with the increase of Ti content $x$ from 0.0 to 1.0, assuming the alloys preserve a bcc phase as highlighted in Figure \ref{fig:lattice}(a). In Figure \ref{fig:lattice}(b), the lattice parameters (a$'$ = (a+b)/2, c/a$'$) for the hcp phase are shown. These properties progressively converge to the ones of pure Ti as Ti content $x$ increases. We recall that the a and c values for hcp Ti are 2.94 and 4.64, respectively. Finally, the DFT calculations reproduce well the experimental lattice parameter for $x$ = 0 and $x$ = 0.2, which gives confidence in our methodology relying on multiple sets of MCSQS and DFT simulations. For detailed lattice parameter analysis, the reader can consult the supplementary Table \ref{tab:S1}.

\begin{figure}[ht]
    \centering
    \includegraphics[scale=0.43]{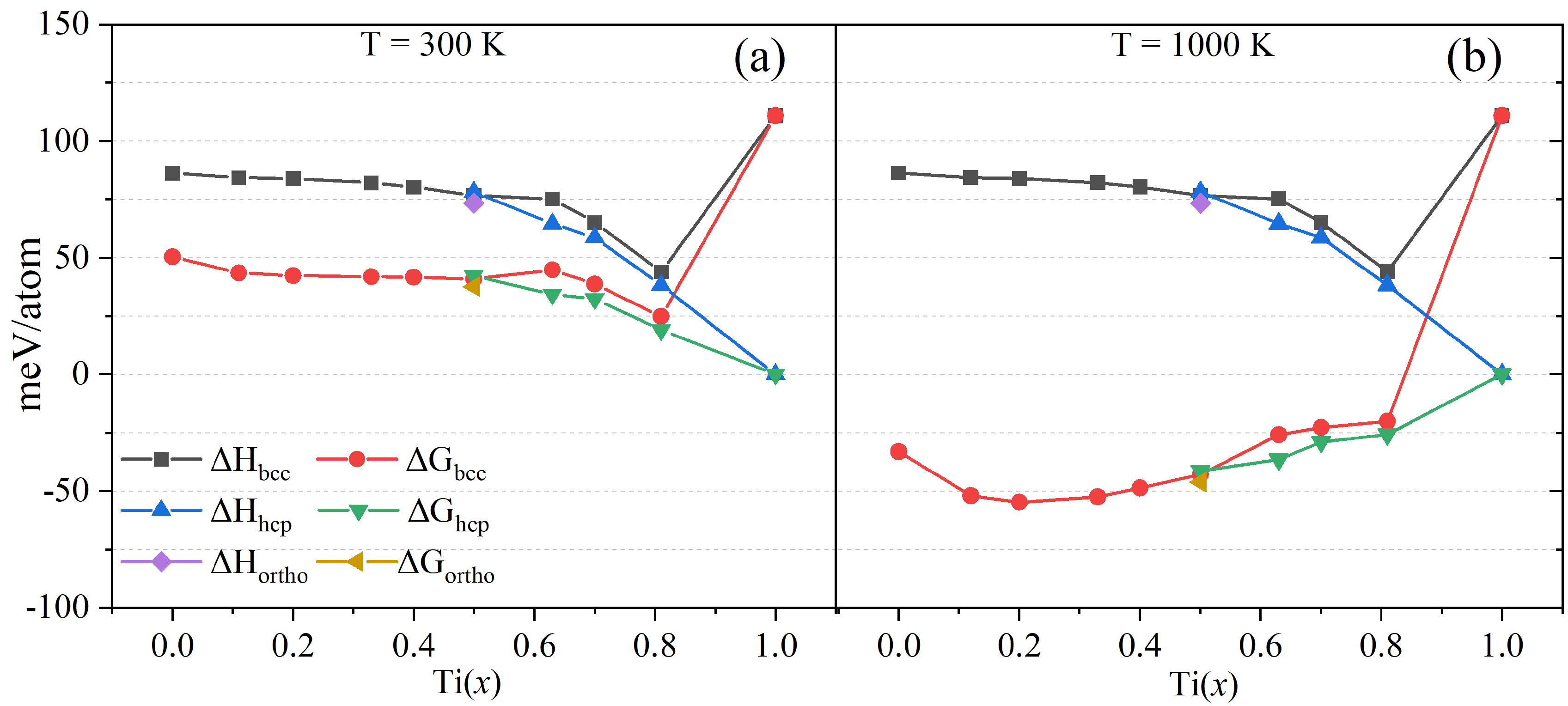}
    \caption{Calculated mixing energy ($\Delta$H) and Gibbs free energy ($\Delta$G) of Ti$_x$(HfNbTaZr)$_{(1-x)/4}$ using the reference state as (a) stable (bcc and hcp phases for bulk systems. In (a), the Gibbs free energy is given at 300 K. Figure (b) is similar to (a), but the Gibbs free energy is now calculated at 1000 K. }
    \label{fig:Gibbs_Energy}
\end{figure}

Figure \ref{fig:Gibbs_Energy}(a) shows the computed mixing energies of the possible phases of Ti content $x$. Mixing energies are calculated using Eq. \ref{eq:Gibbs2} by taking the energy of the most stable phase of each alloying element as a reference (namely, bcc for Nb, hcp for Ti, and so on). Interestingly, the mixing energies of all phases of the alloy compositions are slightly positive for all Ti content $x$. The corresponding values are, however small, below 100 meV/atom.

The mixing energy of the bcc phase decreases monotonically with the Ti content $x$. For $x$ = 0.20, which corresponds to the equimolar case, the mixing energy of 84.4 meV/atom determined in the present study is similar to the DFT calculations of Gao \textit{et al.} \cite{Gao2016a}, where a value of 86.5 meV/atom was reported. %Which is related to the exploration of a large configuration space and finding a corresponding structure with the lowest energy.

% AI: Okay, let me check my files; I might have these data. 
%\textcolor{red}{[SQ: Ok, I'm sorry to add one more thing, but what if on the left we take natural structural references (bcc + hcp) with T = 0,300 and 1000K and on the right forced bcc structural references and curves for T = 0,300 and 1000K ?], => SQ: I saw your answer somewhere else if using the unnatural bcc structure for all element is not done in the community, maybe it is best not to do it and keep the current figure AI: Then I will keep the figure as it is. }

If we compare the relative stability of the possible phases for $x > 0.4$ in Figure \ref{fig:Gibbs_Energy}(a), mixing energy for the hcp phase becomes lower than that of the bcc phase and decreases faster as Ti content $x$ is increased. This suggests that alloys with a large Ti content may exhibit an hcp structure or dual bcc and hcp phases. The mixing energy difference between the two structures is very low (i.e., 2 meV/atom for $x$ = 0.50 and 5 meV/atom for $x$ = 0.63). The mixing energy of the orthorhombic phase at $x$ = 0.50 is slightly smaller than that of the bcc and hcp phases and could constitute a third possible structure. These phases could well be found in the synthesized unstrained samples or reveal themselves by phase transformation upon straining, like in a related alloy \cite{Lilensten2017a}.

Figure \ref{fig:Gibbs_Energy}(a) also displays the mixing energies of pure Ti in the bcc and hcp phases. The mixing energy of Ti$_x$(HfNbTaZr)$_{(1-x)/4}$ alloys in the hcp structure smoothly decreases towards that of pure Ti in the hcp structure. In the context of high entropy alloys, the mixing entropy term has been thought to stabilize the HEA phase by lowering the Gibbs free energy. The consensus is now more nuanced, as it has been shown that high mixing entropy is neither a sufficient nor a necessary condition for the formation of a single-phase solid solution \cite{Otto2013, Laurent-Brocq2016,Tomilin2015}. Here, the Gibbs free energy calculated at RT accounting for the mixing entropy term as given in Eq. \ref{eq:Gibbs3} is still not enough to reach negative values.

%\textcolor{red}{[SQ: I thought formation energy was when taking the 'natural' reference structure energies for all elements and mixing energy for the forced 'bcc' structure, isn't it? Check the usage of formation and usage:] \textcolor{teal}{AI: In the HEA, mixing enthalpy/mixing energy term is mostly used \cite{Song2017} Song, H., (2017). Local lattice distortion in high-entropy alloys. Physical Review Materials, 1(2), 23404. https://doi.org/10.1103/PhysRevMaterials.1.023404. As I read, people in the physics/DFT community preferred formation energy and chemistry enthalpy, and to my understanding, these two concepts are the same. The choice of reference is important, and in our case, we used bulk phase rather than pure single atoms as a stable criterion. Song2017 and many others have used this definition. Previously, when I used formation, I used it from the mindset of the DFT community.} 
%\textcolor{blue}{SQ: ok}
%Also, I am confused about Fig 10.b don't we use the bcc structure for all reference energies for all data points? In the text, it seems that this is only the case for Gibbs energy. [AI: That was an old figure where I used all the bcc reference energies, and then David told me to change the figure with stable reference energies.]}
%\textcolor{blue}{SQ: so that is why you only use the all bcc reference structure at high temperature?}

The fact that the formation energy remains positive, albeit slightly, while a stable solid solution phase is observed for at least some of the alloy compositions in the experiments can be explained in several ways. First, the solid solution mixing energy is strongly impacted by the structure reference chosen for the elements, most of which are stable in hcp at 0K while the others are stable in bcc structure. These HEA are, however, typically elaborated at temperatures well above 1000K, and all the elements in presence exhibit typically a bcc structure at high temperatures.

%\textcolor{red}{ The fact that the formation energy remains positive, albeit slightly, while a stable solid solution phase is observed for at least some of the alloy compositions in the experiments can be explained in several ways. First, the solid solution mixing energy is strongly impacted by the structure reference chosen for the elements, most of which are stable in hcp at 0K while the others are stable in bcc structure. These HEA are, however, typically elaborated at temperatures well above 1000K, and all the elements in presence exhibit typically a bcc structure at high temperatures [SQ: Check].} \textcolor{teal}{[AI: For this, I need to plot again because the current figure just takes the stable reference energies. David told me to report at 1000 K but we are talking about at room temperature, where we experimentally observe a stable phase. Should we not talk about it at room temperature? All DFT results are applicable at 0K.}

%\textcolor{red}{ If we now consider the bcc structure as a reference for all the elements (see Figure \ref{fig:Gibbs_Energy}(b)), the Gibbs free energy becomes negative for all considered alloy compositions. Besides, when considering high temperatures, vibrational entropy S could also play a stabilizing role. Interestingly, the quinary equimolar alloy ($x = 0.20$) exhibits the lowest mixing energy of all the considered compositions, mostly due to a larger contribution from entropy.}

If we now consider the bcc structure as a reference for all the elements (see Figure \ref{fig:Gibbs_Energy}(b)), the Gibbs free energy becomes negative for all considered alloy compositions. Besides, when considering high temperatures, vibrational entropy S could also play a stabilizing role. Interestingly, the quinary equimolar alloy ($x = 0.20$) exhibits the lowest mixing energy of all the considered compositions, mostly due to a larger contribution from entropy.

%%%%%%%%%%%%%%%%%%%%%%%%%%%%%%%%%%%%%%%%%%%%%%%%%%%%%%%%%%
\section{Elastic constants for bcc phase}

In this final section, we evaluate the elastic constants of the bcc phase as function of the Ti content. The energy-strain approach is a commonly used method to determine the elastic constants at 0 K. In this study, the strain tensor was considered based on the symmetry of the relaxed supercell, as described in \cite{Jochym2000}. To accurately determine the elastic constants, a strain range of +/-2$\%$ was applied to the equilibrium structure, resulting in 11 deformed structures. These structures were then relaxed to obtain the atomic positions. To focus specifically on the bcc elastic properties, calculations were limited to alloys with a Ti content $x < 0.5$, where the bcc phase is known to be the most stable. 

\begin{figure}[ht]
    \centering
    \includegraphics[scale=0.31]{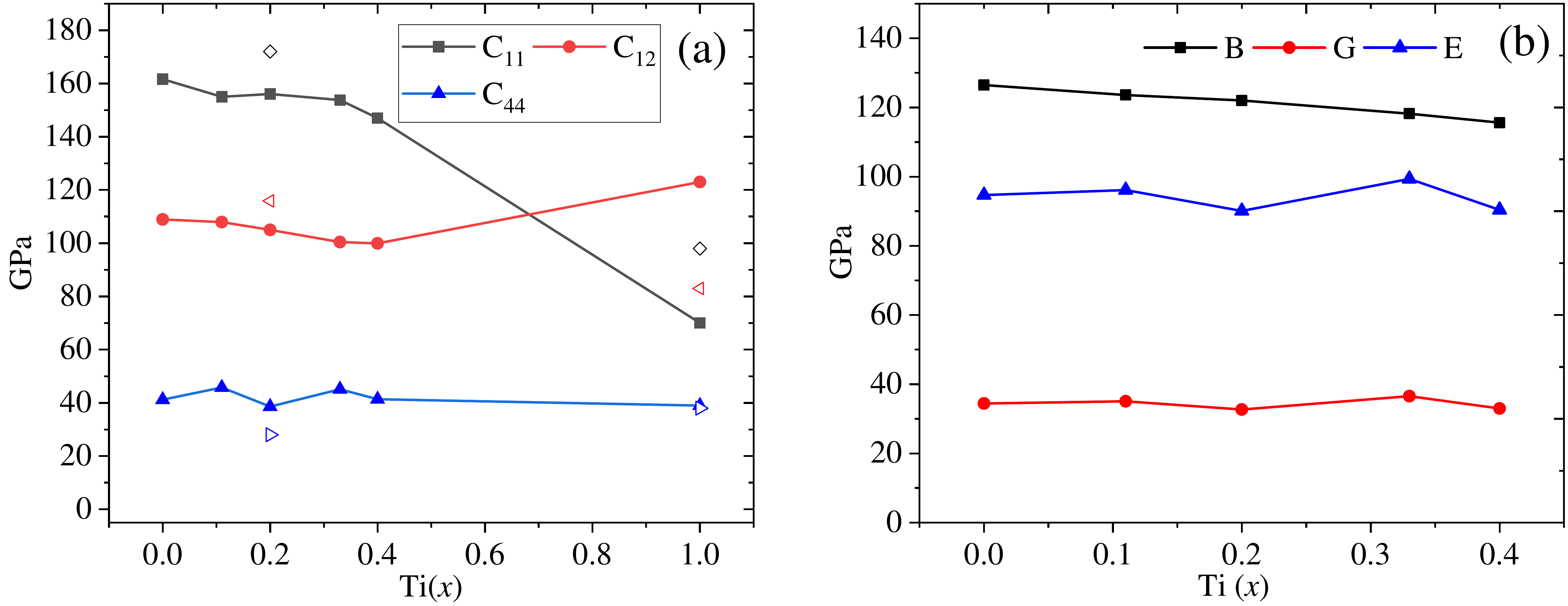}
    \caption{Calculated elastic moduli for the bcc SQS supercell for Ti$_x$(HfNbTaZr)$_{(1-x)/4}$ as a function of the Ti content $x$. (a) the elastic constants $C_{ij}$ (GPa) and (b) the isotropic averages. The elastic constants of bcc Ti are also given for comparison. The experimental values are shown with hollow symbols \cite{Dirras2016, Ledbetter2004}. Corresponding data can be found in supplementary Table \ref{tab:S2}.} 
    \label{fig:Elastic_constants}
\end{figure}

The calculated average elastic constants of the alloy Ti$_x$(HfNbTaZr)$_{(1-x)/4}$ ($x$ = 0.0 to 0.4) are given in Figure \ref{fig:Elastic_constants}(b). These calculations were carried out for the atomic configuration corresponding to the lowest mixing energy for each $x$. Figure \ref{fig:Elastic_constants}(b) shows that $C_{11}$ decreases slowly at first and then rapidly with the increase of Ti content, while $C_{12}$ decreases at first and $C_{44}$ remains mostly unchanged.  Stability conditions on elastic constants also provide an alternative way of assessing the structural stability of alloys.  We therefore checked that the elastic constants for $x$ from 0 to 0.4 satisfy the following conditions, i.e., $C_{11} > C_{12}$, $C_{11} + 2C_{12} > 0$, and $C_{44} > 0$. These conditions are, however, not fulfilled for large Ti content $x$ > 0.5 (when extrapolating the simulation results past $x$ > 0.5). This loss of mechanical stability of the bcc phase is thus in agreement with the Bo-Md diagram prediction and our results regarding the mixing energies of the bcc and hcp phases in the previous section. Finally, elastic anisotropy can be discussed through the Zener anisotropy ratio $A_{z}$. Anisotropy starts low for the quaternary alloy $A_{z}$ = 1.5, and the Zener coefficient then increases from 1.5 to 1.9 when the Ti content is increased from $x$ = 0.2 to 0.4.

A detailed comparison can be made between the simulated and experimental data for an equimolar quinary alloy, where we can see a good agreement \cite{Dirras2016, Ledbetter2004}. The largest difference observed is around 15\% for the elastic constant $C_{11}$. However, when examining pure bcc Ti, the agreement between simulation and experiment is not as satisfactory. A comparison can also be made with another simulation concerning equimolar quinary alloys carried out using the coherent potential approximation (CPA) \cite{Fazakas2014}. It yielded $C_{11}$ = 160.2 GPa, $C_{12}$ = 124.4 GPa, and $C_{44}$ = 62.4 GPa. While the first two constants correspond well to experimental constants, the last value seems particularly high compared to experimental data and our own. The coherent potential approximation assumes a perfect mixing of atoms within the alloy, neglecting any local atomic ordering or clustering that may occur. On the other hand, our simulation accounts for these effects by using a more realistic atomic model. This could explain the higher value obtained for $C_{44}$ in the equimolar quinary alloy. 

% \textcolor{red}{[SQ: I like the new figure; I would put it on the left and the isotropic average on the right then. Also, this is strange to see the Az values as they are close to 1 on the same graph; I would put the Az values in a table somewhere, maybe along with all other elastic constant values in the appendixes. AI: I will put that in the appendix. => ok thanks!}}

The isotropic bulk modulus $B$, shear modulus $G$, and Young’s modulus $E$ can be evaluated from previous elastic constants for the various Ti content $x$. The shear modulus was obtained from the average of the Voigt and Reuss expressions, as suggested by Hill \cite{Hill1952}. Figure \ref{fig:Elastic_constants}(b) shows the evolution of the isotropic elastic constants as a function of the Ti content. $B$ decreases monotonously when increasing the Ti content, while $G$ and $E$ are roughly constant for $x < 0.5$. Finally, Pugh's ratio $(B/G)$ indicates the brittle or ductile nature of the material's behavior \cite{Pugh1954}. Materials that present a large $B$ and a low $G$ typically tend to exhibit ductile behavior. Mechanical testing conducted on the $x$ = 0.2 \cite{Dirras2016} already demonstrated the ductile behavior of this composition. Based on our calculations, Pugh’s ratio $B/G$ is $>$ 1.75 for all Ti content $x$ from 0 to 0.4, suggesting a ductile behavior for this entire range of compositions. %Ultimately, the nice agreement between our results and experimental and simulation works again provides confidence in our approach and calculations based on a large number of MCSQS realizations prior to the DFT simulations.
%\textcolor{red}{[SQ: what is the other study on the quaternary alloy?]} AI: There is a mechanical study on the quaternary alloy. SQ: is there any Elastic constant evaluation? AI: I mean quinary, not quaternary. I could not find any study on quaternary alloys.

%%%%%%%%%%%%%%%%%%%%%%%%%%%%%%%%%%%%%%%%%%%%%%%%%%%%%%%%%%%%%%%%%%%%%%%%%%%%%%%%%%%
\section{Conclusions}

In this work, we have studied the structure, phase stability, and elastic behavior of \\ Ti$_x$(HfNbTaZr)$_{(1-x)/4}$ alloys as a function of the Ti content $x$. The atomistic structure, corresponding to the disordered structure of the HEA solid solution, was approximated using a large number of different MCSQS realizations, while the atomic relaxation and energy calculations were performed using DFT calculations for each MCSQS configuration. The following results were obtained:

\begin{itemize}
\item We conducted a systematic sensitivity study on the number of NNs and on the order of the many-body terms employed to approximate the disordered atomic configuration in the MCSQS. In agreement with other studies, we found that the number of NN pairs has a stronger effect than higher bond orders to obtain SQS configurations corresponding to lower mixing energies.

\item We show that the ideal solid solution is better approximated by considering several MCSQS realizations of the atomic disordered structure. The average mixing energy and global Short Range Order (SRO) parameter converge, however, faster than the SRO parameters for each bond type. The averages over multiple MCSQS configurations will converge faster toward the ideal solid solution representation for simpler alloy compositions and/or bigger supercells. 

    %\textcolor{red}{
    %\item Adjusting the Ti content $x$, which corresponds to the number of atoms and the size of the supercell, affects the mixing energy of the system. An optimum of 8 NN p-shells is found for $x$ = 0.0, 0.20, and $x$ = 0.33. For $x$ = 0.12 and 0.50, 6 NN was sufficient. This shows that depending on the composition and the initial (p, t, q) NN condition, it is necessary to sample enough SQS structures to obtain a reliable, stable structure.}

    \item The mixing energy of the alloys in the bcc structure is slightly positive; the value is, however, smaller than the absolute value of the bulk energy of individual elements. The sign of Gibbs free energy depends on the choice of structure reference for the various elements. A negative Gibbs free energy is obtained when considering the bcc phase as the reference phase for all elements or when including the configuration entropy at the high elaboration temperature. The bcc HEA phase observed experimentally for the equimolar quinary alloy could thus be metastable or stabilized thanks to SRO and/or vibrational entropy.

    \item We have proposed an original analysis of the short-range chemical ordering in the HfNbTaZr and Ti(HfNbTaZr) systems by making use of the large number of MCSQS realizations and DFT evaluations of the corresponding energies. The different configurations can be sorted as a function of the associated mixing energy, and correlations can be pursued. Our SRO analysis strategy can be employed to quickly separate very favorable from unfavorable atomic pairs and is particularly efficient for simpler systems (i.e., quaternary alloys).

    \item We found that Ti-Zr, Ti-Hf, and Nb-Ta are particularly favored at short and longer distances (1 and 3 NN), probably leading to local fluctuations of the composition in the alloy. This could explain the waviness of dislocations and debris observed in experiments. Our SRO analysis is also mostly in agreement with the few existing studies relying on a more advanced MC method.

    \item To simplify the complex SRO analysis, elements with identical crystallographic structures are typically favorable, while elements with different crystallographic structures are unfavorable, and their presence is minimized in the LE SQS sets. It would be interesting to see if this simple explanation can be applied to other HEA systems with elements of different crystallographic structures. The SRO in HfNbTaZr is mainly located on the 2nd NN shell, while in HfNbTaTiZr all the first three NN shells are involved. Few bond pairs are favorable in the quaternary system, and this could explain the weak global SRO in this system. Short-range ordering appears to be present in both systems, although it is stronger in the quinary system. 

    \item The content of Ti has various effects on the structure or properties of the \\ Ti$_x$(HfNbTaZr)$_{(1-x)/4}$ system. The average bcc lattice parameter and the mixing energies decrease monotonically with the Ti composition. Ti content can be tailored to adapt the elastic properties to specific applications ($C_{11}$ and $B$) by a few tens of GPa. The Ti content also controls the structure of the HEA phase, with a bcc phase stable, predicted as ductile, for $x$ < 0.5. For Ti content larger than 0.5, the hcp phase becomes the most stable. Close to the quinary equimolar composition, a dual bcc-hcp phase or deformation induced phase transformation may be possible. A third possible phase in the shape of an orthorhombic structure is discarded when the mixing entropy term is considered. These results are consistent with the so-called Bo-Md diagram, which was originally proposed to predict the phase stability of Ti alloys.

    \item The present work is qualitatively consistent with trends found in HEA systems and quantitatively consistent with the limited literature on this system. The average bcc lattice parameter and elastic constants agree with the experimental values obtained for the equimolar alloy. The mixing energy is also in agreement with the results obtained for the equimolar alloy in \cite{Zhang2021,Gao2016a,Xun2023}.

    \item The results regarding dislocation structure and properties that control the plastic behavior and forming capability of this system are left for a forthcoming paper [Asif I. Bhatti \textit{et al.} to be published 2024]
\end{itemize}

\section{Acknowledgements}
We are grateful to the University Sorbonne, Paris Nord, Labex SEAM, ANR (Agence Nationale de la Recherche), and CGI (Commissariat à l’Investissement d’Avenir) for financing this work through Labex SEAM (Science and Engineering for Advanced Materials and devices), ANR-10-LABX-0096 and ANR-18-IDEX-0001. Further, we would like to thank Ing. Nicolas Greneche for HPC support, MAGI, Univ. Sorbonne Paris Nord, and HPC resources. We would like to thank Guy Dirras and Céline Varvenne for their helpful suggestions and discussions that improved the manuscript.

%%%%%%%%%%%%%%%%%%%%%%%%%%%%%%%%%%%%
\clearpage
\bibliographystyle{unsrt}
\bibliography{main.bib}

\begin{thebibliography}{10}

\bibitem{Gao}
M~C Gao, C~S Carney, N.~Doğan, P~D Jablonksi, J~A Hawk, and D~E Alman.
\newblock Design of refractory high-entropy alloys.
\newblock {\em JOM}, 67:2653--2669, 2015.

\bibitem{Miracle2017a}
D.~B. Miracle and O.~N. Senkov.
\newblock A critical review of high entropy alloys and related concepts, 1 2017.

\bibitem{Ikeda2019a}
Yuji Ikeda, Blazej Grabowski, and Fritz Körmann.
\newblock Ab initio phase stabilities and mechanical properties of multicomponent alloys: A comprehensive review for high entropy alloys and compositionally complex alloys, 1 2019.

\bibitem{Dirras2016}
G.~Dirras, L.~Lilensten, P.~Djemia, M.~Laurent-Brocq, D.~Tingaud, J.~P. Couzinié, L.~Perrière, T.~Chauveau, and I.~Guillot.
\newblock Elastic and plastic properties of as-cast equimolar tihfzrtanb high-entropy alloy.
\newblock {\em Materials Science and Engineering A}, 654:30--38, 1 2016.

\bibitem{Senkov2018}
O.~N. Senkov, S.~Rao, K.~J. Chaput, and C.~Woodward.
\newblock Compositional effect on microstructure and properties of nbtizr-based complex concentrated alloys.
\newblock {\em Acta Materialia}, 151:201--215, 6 2018.

\bibitem{Gao2016c}
Michael~C. Gao, Changning Niu, Chao Jiang, and Douglas~L. Irving.
\newblock Applications of special quasi-random structures to high-entropy alloys.
\newblock {\em High-Entropy Alloys: Fundamentals and Applications}, pages 333--368, 1 2016.

\bibitem{Takeuchi2013}
Akira Takeuchi, Kenji Amiya, Takeshi Wada, Kunio Yubuta, Wei Zhang, and Akihiro Makino.
\newblock Entropies in alloy design for high-entropy and bulk glassy alloys.
\newblock {\em Entropy}, 15:3810--3821, 2013.

\bibitem{Yao2017}
H.~W. Yao, J.~W. Qiao, J.~A. Hawk, H.~F. Zhou, M.~W. Chen, and M.~C. Gao.
\newblock Mechanical properties of refractory high-entropy alloys: Experiments and modeling.
\newblock {\em Journal of Alloys and Compounds}, 696:1139--1150, 2017.

\bibitem{Manzoni2020}
Anna~M. Manzoni and Uwe Glatzel.
\newblock High-entropy alloys: Balancing strength and ductility at room temperature.
\newblock {\em Reference Module in Materials Science and Materials Engineering}, 1 2020.

\bibitem{Yuan2019}
Yuan Yuan, Yuan Wu, Zhi Yang, Xue Liang, Zhifeng Lei, Hailong Huang, Hui Wang, Xiongjun Liu, Ke~An, Wei Wu, and Zhaoping Lu.
\newblock Formation, structure and properties of biocompatible tizrhfnbta high-entropy alloys.
\newblock {\em Materials Research Letters}, 7:225--231, 6 2019.

\bibitem{Gurel2020}
S.~Gurel, M.~B. Yagci, B.~Bal, and D.~Canadinc.
\newblock Corrosion behavior of novel titanium-based high entropy alloys designed for medical implants.
\newblock {\em Materials Chemistry and Physics}, 254:123377, 11 2020.

\bibitem{Abdel-Hady2006}
Mohamed Abdel-Hady, Keita Hinoshita, and Masahiko Morinaga.
\newblock General approach to phase stability and elastic properties of $\beta$-type ti-alloys using electronic parameters.
\newblock {\em Scripta Materialia}, 55:477--480, 9 2006.

\bibitem{Lilensten2017a}
Lola Lilensten, Jean~Philippe Couzinié, Julie Bourgon, Loïc Perrière, Guy Dirras, Frédéric Prima, and Ivan Guillot.
\newblock Design and tensile properties of a bcc ti-rich high-entropy alloy with transformation-induced plasticity.
\newblock {\em Materials Research Letters}, 5:110--116, 3 2017.

\bibitem{Lei2018}
Zhifeng Lei, Xiongjun Liu, Yuan Wu, Hui Wang, Suihe Jiang, Shudao Wang, Xidong Hui, Yidong Wu, Baptiste Gault, Paraskevas Kontis, Dierk Raabe, Lin Gu, Qinghua Zhang, Houwen Chen, Hongtao Wang, Jiabin Liu, Ke~An, Qiaoshi Zeng, Tai~Gang Nieh, and Zhaoping Lu.
\newblock Enhanced strength and ductility in a high-entropy alloy via ordered oxygen complexes.
\newblock {\em Nature}, 563:546--550, 11 2018.

\bibitem{Bu2021}
Yeqiang Bu, Yuan Wu, Zhifeng Lei, Xiaoyuan Yuan, Honghui Wu, Xiaobin Feng, Jiabin Liu, Jun Ding, Yang Lu, Hongtao Wang, Zhaoping Lu, and Wei Yang.
\newblock Local chemical fluctuation mediated ductility in body-centered-cubic high-entropy alloys.
\newblock {\em Materials Today}, 46:28--34, 6 2021.

\bibitem{Zhang2020}
Ruopeng Zhang, Shiteng Zhao, Jun Ding, Yan Chong, Tao Jia, Colin Ophus, Mark Asta, Robert~O. Ritchie, and Andrew~M. Minor.
\newblock Short-range order and its impact on the crconi medium-entropy alloy.
\newblock {\em Nature}, 581:283--287, 5 2020.

\bibitem{Yina}
Sheng Yin, Jun Ding, Mark Asta, and Robert~O Ritchie.
\newblock Ab initio modeling of the energy landscape for screw dislocations in body-centered cubic high-entropy alloys.
\newblock {\em npj Computational Materials}, 6, 2020.

\bibitem{Xun2023}
Kaihui Xun, Bozhao Zhang, Qi~Wang, Zhen Zhang, Jun Ding, and En~Ma.
\newblock Local chemical inhomogeneities in tizrnb-based refractory high-entropy alloys.
\newblock {\em Journal of Materials Science \& Technology}, 135:221--230, 2 2023.

\bibitem{Chen2021}
Shuai Chen, Zachary~H. Aitken, Subrahmanyam Pattamatta, Zhaoxuan Wu, Zhi~Gen Yu, David~J. Srolovitz, Peter~K. Liaw, and Yong~Wei Zhang.
\newblock Simultaneously enhancing the ultimate strength and ductility of high-entropy alloys via short-range ordering.
\newblock {\em Nature Communications 2021 12:1}, 12:1--11, 8 2021.

\bibitem{Zhang2021}
Bozhao Zhang, Jun Ding, and En~Ma.
\newblock Chemical short-range order in body-centered-cubic tizrhfnb high-entropy alloys.
\newblock {\em Applied Physics Letters}, 119, 11 2021.

\bibitem{Wang2021c}
S.~D. Wang, X.~J. Liu, Z.~F. Lei, D.~Y. Lin, F.~G. Bian, C.~M. Yang, M.~Y. Jiao, Q.~Du, H.~Wang, Y.~Wu, S.~H. Jiang, and Z.~P. Lu.
\newblock Chemical short-range ordering and its strengthening effect in refractory high-entropy alloys.
\newblock {\em Physical Review B}, 103:104107, 3 2021.

\bibitem{Yin2020}
Binglun~; Yin, Francesco~; Maresca, William Curtin, and Acta Materialia.
\newblock Vanadium is an optimal element for strengthening in both fcc and bcc high-entropy alloys.
\newblock 2020.

\bibitem{Yin2019}
Binglun Yin and William~A. Curtin.
\newblock First-principles-based prediction of yield strength in the rhirpdptnicu high-entropy alloy.
\newblock {\em npj Computational Materials}, 5:14, 12 2019.

\bibitem{Varvenne2016a}
Céline Varvenne, Aitor Luque, and William~A. Curtin.
\newblock Theory of strengthening in fcc high entropy alloys.
\newblock {\em Acta Materialia}, 118:164--176, 10 2016.

\bibitem{Lilensten2018}
L.~Lilensten, J.~P. Couzinié, L.~Perrière, A.~Hocini, C.~Keller, G.~Dirras, and I.~Guillot.
\newblock Study of a bcc multi-principal element alloy: Tensile and simple shear properties and underlying deformation mechanisms.
\newblock {\em Acta Materialia}, 142:131--141, 1 2018.

\bibitem{Couzinie2015}
J.~Ph Couzinié, L.~Lilensten, Y.~Champion, G.~Dirras, L.~Perrière, and I.~Guillot.
\newblock On the room temperature deformation mechanisms of a tizrhfnbta refractory high-entropy alloy.
\newblock {\em Materials Science and Engineering: A}, 645:255--263, 10 2015.

\bibitem{Lilensten2022}
Lola Lilensten, Karine Provost, Loïc Perrière, Emiliano Fonda, Jean~Philippe Couzinié, Fabienne Amann, Martin Radtke, Guy Dirras, and Ivan Guillot.
\newblock Experimental investigation of the local environment and lattice distortion in refractory medium entropy alloys.
\newblock {\em Scripta Materialia}, 211:114532, 4 2022.

\bibitem{Tan2023}
Yuan~Yuan Tan, Tong Li, Yu~Chen, Zhong~Jun Chen, Ming~Yao Su, Jing Zhang, Yu~Gong, Tao Wu, Hai~Ying Wang, and Lan~Hong Dai.
\newblock Uncovering heterogeneity of local lattice distortion in tizrhfnbta refractory high entropy alloy by sr-xrd and exafs.
\newblock {\em Scripta Materialia}, 223:115079, 1 2023.

\bibitem{Rao2019}
S~I Rao, B~Akdim, E~Antillon, C~Woodward, T~A Parthasarathy, and O~N Senkov.
\newblock Modeling solution hardening in bcc refractory complex concentrated alloys: Nbtizr, nb1.5tizr0.5 and nb0.5tizr1.5.
\newblock {\em Acta Materialia}, 168:222--236, 2019.

\bibitem{Huang2021}
Xiusong Huang, Lehua Liu, Xianbao Duan, Weibing Liao, Jianjun Huang, Huibin Sun, and Chunyan Yu.
\newblock Atomistic simulation of chemical short-range order in hfnbtazr high entropy alloy based on a newly-developed interatomic potential.
\newblock {\em Materials \& Design}, 202:109560, 4 2021.

\bibitem{Kresse1996}
G.~Kresse and J.~Furthmüller.
\newblock Efficient iterative schemes for ab initio total-energy calculations using a plane-wave basis set.
\newblock {\em Physical Review B}, 54:11169--11186, 10 1996.

\bibitem{Kresse1999}
G.~Kresse and D.~Joubert.
\newblock From ultrasoft pseudopotentials to the projector augmented-wave method.
\newblock {\em Physical Review B}, 59:1758--1775, 1 1999.

\bibitem{Perdew1996}
John~P. Perdew, Kieron Burke, and Matthias Ernzerhof.
\newblock Generalized gradient approximation made simple.
\newblock {\em Physical Review Letters}, 77:3865--3868, 10 1996.

\bibitem{Glass2006}
Colin~W. Glass, Artem~R. Oganov, and Nikolaus Hansen.
\newblock Uspex-evolutionary crystal structure prediction.
\newblock {\em Computer Physics Communications}, 2006.

\bibitem{Soven1967}
Paul Soven.
\newblock Coherent-potential model of substitutional disordered alloys.
\newblock {\em Physical Review}, 156:809--813, 4 1967.

\bibitem{Gao2017b}
Michael~C. Gao, Pan Gao, Jeffrey~A. Hawk, Lizhi Ouyang, David~E. Alman, and Mike Widom.
\newblock Computational modeling of high-entropy alloys: Structures, thermodynamics and elasticity, 10 2017.

\bibitem{Song2017}
Hongquan Song, Fuyang Tian, Qing~Miao Hu, Levente Vitos, Yandong Wang, Jiang Shen, and Nanxian Chen.
\newblock Local lattice distortion in high-entropy alloys.
\newblock {\em Physical Review Materials}, 1:23404, 2017.

\bibitem{Carlsson1986}
A.~E. Carlsson and P.~A. Fedders.
\newblock Maximum-entropy method for electronic properties of alloys.
\newblock {\em Physical Review B}, 34:3567--3571, 9 1986.

\bibitem{VandeWalle2013}
A.~van~de Walle, P~Tiwary, M.~de~Jong, D.L. Olmsted, M~Asta, A~Dick, D~Shin, Y~Wang, L.-Q. Chen, and Z.-K. Liu.
\newblock Efficient stochastic generation of special quasirandom structures.
\newblock {\em Calphad}, 42:13--18, 9 2013.

\bibitem{VanDeWalle2009}
Axel van~de Walle.
\newblock Multicomponent multisublattice alloys, nonconfigurational entropy and other additions to the alloy theoretic automated toolkit.
\newblock {\em Calphad}, 33:266--278, 6 2009.

\bibitem{Methfessel1989}
M.~Methfessel and A.~T. Paxton.
\newblock High-precision sampling for brillouin-zone integration in metals.
\newblock {\em Physical Review B}, 40:3616--3621, 8 1989.

\bibitem{Alfe2009}
Dario Alfè.
\newblock Phon: A program to calculate phonons using the small displacement method.
\newblock {\em Computer Physics Communications}, 180:2622--2633, 12 2009.

\bibitem{Togo2015}
Atsushi Togo and Isao Tanaka.
\newblock First principles phonon calculations in materials science.
\newblock {\em Scripta Materialia}, 108:1--5, 11 2015.

\bibitem{Du2019}
Zhenyu Du, Jie Zuo, Nanyun Bao, Mingli Yang, Gang Jiang, and Li~Zhang.
\newblock Effect of ta addition on the structural, thermodynamic and mechanical properties of cocrfeni high entropy alloys.
\newblock {\em RSC Advances}, 9:16447--16454, 5 2019.

\bibitem{Sobieraj2020}
Damian Sobieraj, Jan~S. Wróbel, Tomasz Rygier, Krzysztof~J. Kurzydłowski, Osman~El Atwani, Arun Devaraj, Enrique~Martinez Saez, and Duc Nguyen-Manh.
\newblock Chemical short-range order in derivative cr-ta-ti-v-w high entropy alloys from the first-principles thermodynamic study.
\newblock {\em Physical Chemistry Chemical Physics}, 22:23929--23951, 11 2020.

\bibitem{Tian2020a}
Fuyang Tian, De~Ye Lin, Xingyu Gao, Ya~Fan Zhao, and Hai~Feng Song.
\newblock A structural modeling approach to solid solutions based on the similar atomic environment.
\newblock {\em Journal of Chemical Physics}, 153, 2020.

\bibitem{DeFontaine1971}
D.~de~Fontaine.
\newblock The number of independent pair-correlation functions in multicomponent systems.
\newblock {\em Journal of Applied Crystallography}, 4:15--19, 1971.

\bibitem{Menon2019}
Sarath Menon, Grisell~Díaz Leines, and Jutta Rogal.
\newblock pyscal: A python module for structural analysis of atomic environments.
\newblock {\em Journal of Open Source Software}, 4:1824, 11 2019.

\bibitem{Yin2021}
Sheng Yin, Yunxing Zuo, Anas Abu-Odeh, Hui Zheng, Xiang~Guo Li, Jun Ding, Shyue~Ping Ong, Mark Asta, and Robert~O. Ritchie.
\newblock Atomistic simulations of dislocation mobility in refractory high-entropy alloys and the effect of chemical short-range order.
\newblock {\em Nature Communications}, 12, 12 2021.

\bibitem{Morinaga1988}
M.~Morinaga, N.~Yukawa, T.~Maya, K.~Sone, and H.~Adachi.
\newblock Theoretical design of titanium alloys, 1988.

\bibitem{Senkov2011}
O.~N. Senkov, J.~M. Scott, S.~V. Senkova, D.~B. Miracle, and C.~F. Woodward.
\newblock Microstructure and room temperature properties of a high-entropy tanbhfzrti alloy.
\newblock {\em Journal of Alloys and Compounds}, 509:6043--6048, 5 2011.

\bibitem{Gao2016a}
Michael~C. Gao, Peter~K. Liaw, Jien~Wei Yeh, and Yong Zhang.
\newblock {\em High-entropy alloys: Fundamentals and applications}.
\newblock Springer International Publishing, 1 2016.

\bibitem{Otto2013}
F.~Otto, Y.~Yang, H.~Bei, and E.~P. George.
\newblock Relative effects of enthalpy and entropy on the phase stability of equiatomic high-entropy alloys.
\newblock {\em Acta Materialia}, 61:2628--2638, 4 2013.

\bibitem{Laurent-Brocq2016}
Mathilde Laurent-Brocq, Loïc Perrière, Rémy Pirès, and Yannick Champion.
\newblock From high entropy alloys to diluted multi-component alloys: Range of existence of a solid-solution.
\newblock {\em Materials and Design}, 103:84--89, 8 2016.

\bibitem{Tomilin2015}
I.~A. Tomilin and S.~D. Kaloshkin.
\newblock 'high entropy alloys'-'semi-impossible' regular solid solutions?
\newblock {\em Materials Science and Technology (United Kingdom)}, 31:1231--1234, 7 2015.

\bibitem{Jochym2000}
P.~T. Jochym and K.~Parlinski.
\newblock Ab initio lattice dynamics and elastic constants of zrc.
\newblock {\em European Physical Journal B}, 15:265--268, 5 2000.

\bibitem{Ledbetter2004}
Massel Ledbetter, Hirotsugu Ogi, Satoshi Kai, Sudook Kim, and Masahiko Hirao.
\newblock Elastic constants of body-centered-cubic titanium monocrystals.
\newblock {\em Journal of Applied Physics}, 95:4642--4644, 5 2004.

\bibitem{Fazakas2014}
E.~Fazakas, V.~Zadorozhnyy, L.~K. Varga, A.~Inoue, D.~V. Louzguine-Luzgin, Fuyang Tian, and L.~Vitos.
\newblock Experimental and theoretical study of ti20zr20hf20nb20x20 (x = v or cr) refractory high-entropy alloys.
\newblock {\em International Journal of Refractory Metals and Hard Materials}, 47:131--138, 11 2014.

\bibitem{Hill1952}
R.~Hill.
\newblock The elastic behaviour of a crystalline aggregate.
\newblock {\em Proceedings of the Physical Society. Section A}, 65:349--354, 5 1952.

\bibitem{Pugh1954}
S.F. Pugh.
\newblock Xcii. relations between the elastic moduli and the plastic properties of polycrystalline pure metals.
\newblock {\em The London, Edinburgh, and Dublin Philosophical Magazine and Journal of Science}, 45:823--843, 8 1954.

\end{thebibliography}

%%%%%%%%%%%%%%%%%%%%%%%%%%%%%%%%%%%%%%%%%%%%%%%%%%%%%%%%%%%%%%%%%%%%%%%%%%%%%%%%%%%
\clearpage
\section*{Supplementary Information}

Table \ref{tab:S1} shows the structural parameters for the configurations corresponding to the lowest mixing energy structures. The relaxed lattice parameters (a, b, c) deviate slightly from the perfect bcc lattice. They are slightly anisotropic (with an average difference of about 3$\%$), and the angles $(\alpha,\beta,\gamma)$ between the basis vectors of the supercell slightly differ from the expected 90$\degree$ (by typically less than 1$\degree$). These small distortions are unsurprising for HEA systems, as differences in the local chemical order are expected along the supercell axis, especially for the small dimensions dealt with in DFT.

The known stable experimental lattice parameters at RT for stable elements are $a_{Ti}$ = 2.95 \AA, $c/a_{Ti}$ = 1.58 \AA, $a_{Zr}$ = 3.23 \AA, $c/a_{Zr}$ = 1.59 \AA, $a_{Hf}$ = 3.196 \AA, $c/a_{Hf}$ = 1.58 \AA, $a_{Nb}$= 3.30 \AA, and $a_{Ta}$ = 3.31 \AA. The lattice constants for a pure Ti (hcp) unit cell are a = b = 2.950 \AA and c = 4.687 \AA. For large Ti content, typically $x > 0.4$ and above, the Bo-Md diagram suggests that hcp and/or orthorhombic (ortho) structures may be stable. We thus performed another set of MCSQS+DFT calculations for these crystallographic structures. We adopted the same methodology as discussed in Section 2. To assess the prediction of the Bo-Md diagram, we generated a set of MCSQS structures, assuming the hcp phase for $x$ ranges from 0.5 to 0.81.

The bottom of Table \ref{tab:S1} (hcp and ortho) shows the lattice geometry obtained for these simulations. The simulated lattices are again slightly distorted when compared to a perfect hcp lattice, certainly because the different element arrangements are probed along the supercell axis. Surprisingly, no clear trend was found regarding the evolution of the c/a ratio as the Ti content increased.

\begin{table}[hp]
    \centering
    \setcounter{table}{0}
    \renewcommand{\thetable}{S\arabic{table}}
    \caption{Detailed analysis of the structure of the atomic configurations after relaxation for Ti$_x$(HfNbTaZr)$_{(1-x)/4}$ alloys as a function of the Ti content $x$ and the choice of crystallographic structure. The average lattice parameters a$'$ are reported for the conventional unit cell. The number of atoms is given for the SQS supercell generated from MCSQS. Known experimental values are reported from \cite{Huang2021,Senkov2011} and are given in parentheses.}
    \label{tab:S1}
    \scalebox{0.9}{
    \begin{tabular}{lcccccccccc}
    \hline
        \toprule
        Phases & $x$ & \#atoms & a [\AA] & b [\AA]& c [\AA]& $\alpha^{\circ}$ & $\beta^{\circ}$ & $\gamma^{\circ}$ & a$'$ [\AA] \\ \hline
        \multirow{10}{*}{bcc}& 0.00 & 128 & 3.454 & 3.417 & 3.438 & 89.6 & 90.0 & 90.0 & 3.436(3.446) \\ 
        ~ & 0.11 & 54 & 3.429 & 3.408 & 3.425 & 89.5 & 89.8 & 89.9 & 3.420 \\ 
        ~ & 0.20 & 125 & 3.398 & 3.417 & 3.405 & 89.7 & 89.6 & 89.8 & 3.407(3.404) \\ 
        ~ & 0.33 & 54 & 3.392 & 3.372 & 3.381 & 89.7 & 90.1 & 89.9 & 3.381 \\ 
        ~ & 0.40 & 54 & 3.371 & 3.336 & 3.403 & 89.5 & 89.9 & 89.6 & 3.370 \\ 
        ~ & 0.50 & 128 & 3.379 & 3.321 & 3.361 & 89.9 & 90.8 & 90.0 & 3.353 \\ 
        ~ & 0.63 & 54 & 3.313 & 3.345 & 3.339 & 90.6 & 88.6 & 90.7 & 3.332 \\ 
        ~ & 0.70 & 54 & 3.300 & 3.305 & 3.350 & 89.2 & 90.4 & 90.5 & 3.318 \\ 
        ~ & 0.81 & 128 & 3.073 & 3.421 & 3.418 & 85.8 & 89.9 & 90.1 & 3.306 \\ 
        ~ & 1.00 & 2 & 3.252 & 3.252 & 3.252 & 90.0 & 90.0 & 90.0 & 3.252 \\ \hline
        ortho & 0.50 & 32 & 3.430 & 4.684 & 4.707 & 86.9 & 88.2 & 88.0 & ~ \\ \hline
        \multirow{5}{*}{hcp}& 0.50 & 128 & 2.933 & 2.942 & 4.680 & 90.1 & 89.7 & 110.9 & c/a$'$ =1.593 \\ 
        ~ & 0.63 & 54 & 3.318 & 2.961 & 4.543 & 89.9 & 90.0 & 123.8 & c/a$'$ =1.447 \\ 
        ~ & 0.70 & 54 & 2.953 & 3.285 & 4.533 & 89.7 & 89.6 & 123.7 & c/a$'$ =1.453 \\ 
        ~ & 0.81 & 128 & 2.934 & 2.997 & 4.687 & 90.0 & 89.9 & 119.3 & c/a$'$ =1.580 \\ 
        ~ & 1.00 & 2 & 2.94 & 2.94 & 4.64 & 90.0 & 90.0 & 120.0 & c/a$'$ = 1.578(1.586) \\
        \bottomrule
    \end{tabular}}
\end{table}

%  AS PER CELINE I HAVE COMMENTED ON THIS TABLE AND INSTEAD DRAW A FIGURE
\begin{table}[ht]
    \centering
    \setcounter{table}{1}
    \renewcommand{\thetable}{S\arabic{table}}
    \caption{Elastic constants $C_{ij}$ (GPa) and Zener anisotropy coefficient $A_{z}$ calculated for the bcc SQS supercell. The elastic constants of bcc Ti are also given for comparison. The experimental value for the known composition is given in parentheses.}
    \label{tab:S2}
    \begin{tabular}{lccccccc}
        \toprule
         $x$& 0.0& 0.12& 0.20 \cite{Dirras2016}& 0.33& 0.40& 1.0\cite{Ledbetter2004} \\
         \midrule
          $C_{11}$&	162&	155&	156(172$\pm$6)	 &   154	&147	&70(98)\\
          $C_{12}$&	109&	108&	105(108$\pm$1.5) &	 100	&100	&123(83)\\
          $C_{44}$&	42 &    46 &   39(28$\pm$1.5)	 &   45  	&41     &39(38)\\
          $A_{z}$	&   1.5&	1.9&	1.5(1.0)	     &   1.7	&1.9	&-\\
         \bottomrule
    \end{tabular}
\end{table}
\end{document}